\documentclass[11pt,a4paper]{article}
\usepackage[T1]{fontenc}
\usepackage[utf8]{inputenc}
\usepackage[margin=27mm]{geometry}
\usepackage{amsmath,amssymb,graphicx}
\usepackage[numbers,sort&compress]{natbib}
\usepackage[hidelinks]{hyperref}
\usepackage{lmodern,microtype,mathtools,mathrsfs}
\usepackage{booktabs,longtable,array,enumitem}
\hypersetup{pdftitle={Compensator-free N=2 extension of ModMax and scalar propagation},pdfauthor={Ariunzul Davgadorj},pdfsubject={Self-dual supersymmetric electrodynamics and scalar propagation}}
\allowdisplaybreaks[2]
\setlist{itemsep=3pt,topsep=4pt}
\newcommand{\keywords}[1]{\par\medskip\noindent\textbf{Keywords: }#1\par}

\newcommand{\W}{\mathcal W}
\newcommand{\bW}{\overline{\mathcal W}}
\newcommand{\M}{\mathcal M}
\newcommand{\hK}{\widehat K}
\newcommand{\dd}{\mathrm d}
\newcommand{\cc}{\mathrm{c.c.}}
\newcommand{\IC}{\mathfrak I_C}
\newcommand{\IF}{\mathfrak I_F}
\newcommand{\res}{\operatorname{Res}_1}
\newcommand{\im}{\operatorname{Im}}

\newcommand{\AS}{\mathcal A_S}

\newcommand{\SM}{S_{\mathrm M}}
\newcommand{\RB}{R_{\mathrm B}}
\newcommand{\Rr}{R_{\mathrm r}}
\newcommand{\UU}{\mathcal U}
\newcommand{\HH}{\mathscr H}
\newcommand{\PP}{\mathcal P}
\newcommand{\FF}{\mathcal F}
\newcommand{\rhoh}{\rho_\eta}
\newcommand{\FI}[1]{\int\dd^4x\,\dd^2\theta\,\dd^2\bar\theta\,#1}
\newcommand{\CI}[1]{\int\dd^4x\,\dd^2\theta\,#1}
\title{Compensator-free $\mathcal N=2$ extension of ModMax\\and scalar propagation}
\author{Ariunzul Davgadorj\\[0.6em]
\small Department of Mathematics and Physics, Kielce University of Technology\\
\small al. Tysi\k{a}clecia Pa\'nstwa Polskiego 7, 25-314 Kielce, Poland\\[0.4em]
\small\href{mailto:ariunzul.d@gmail.com}{\texttt{ariunzul.d@gmail.com}}}
\date{}

\begin{document}
\maketitle
\begin{abstract}
We construct a compensator-free $\mathcal N=2$ superconformal and
electric--magnetic duality-invariant extension of ModMax electrodynamics
in ordinary flat superspace. The action is built entirely from the
Abelian vector multiplet, using a derivative composite of its field
strength, and the self-duality equation is solved to all orders as a
formal series in the coupling. Upon setting the additional
$\mathcal N=1$ chiral multiplet to zero, the action is equivalent off
shell to the $\mathcal N=1$ ModMax action of Bandos, Lechner, Sorokin
and Townsend through a local vector-prepotential redefinition.

Since the complex scalar of the $\mathcal N=2$ multiplet is set to zero in this reduction, we add one superconformal interaction that leaves the reduction unchanged and is chosen to ensure causal scalar propagation.
On constant non-null electromagnetic backgrounds with vanishing
scalar expectation value, the higher-order scalar coefficients are
fixed by requiring the scalar quadratic action to satisfy the
stress-tensor square-root flow, modulo equations of motion and a
total derivative. We determine the quadratic action of the full
multiplet on these backgrounds to all orders. Its analytic
continuation to finite $\gamma\geq0$ gives causal scalar propagation,
with transverse speed squared $e^{-\gamma}$ and unit longitudinal
speed, the ModMax photon cones, luminal gauginos with positive kinetic
coefficients, and auxiliary fields with algebraic equations of motion.
\end{abstract}


\clearpage
\tableofcontents
\clearpage

\section{Introduction}
\label{sec:intro}
The class of $U(1)$ duality-invariant theories extends well
beyond Maxwell and Born--Infeld theory~\cite{GR1995}; reviews of the
bosonic and supersymmetric formulations are given
in~\cite{KTReview2001,Sorokin2022}. ModMax is a one-parameter nonlinear
extension of four-dimensional Maxwell electrodynamics that preserves
both conformal invariance and $U(1)$ duality invariance~\cite{BLST2020}.
With metric $(-,+,+,+)$, its Lagrangian is
\begin{equation}
 x=-\frac14F_{\mu\nu}F^{\mu\nu},\qquad
 p=-\frac14F_{\mu\nu}\widetilde F^{\mu\nu},\qquad
 r=\sqrt{x^2+p^2},\qquad
 L_{\rm MM}=x\cosh\gamma+r\sinh\gamma.
 \label{eq:bosmodmax}
\end{equation}
The square root is nonanalytic at $r=0$. Accordingly,
our construction is a power series in the coupling on the domain
$r>0$, rather than a Taylor expansion in the field strength about the
vacuum. The coupling of ModMax to charged particles was studied
in~\cite{LMSS2023}.

Supersymmetric Born--Infeld theory relates nonlinear electrodynamics to
partial supersymmetry breaking~\cite{BG1997}; manifestly $\mathcal N=2$
actions and their iterative extensions were studied in
\cite{Ketov1999,BIK2001,BKSS2013}. Ivanov and Zupnik gave an auxiliary-superfield formulation of the
$\mathcal N=2$ Born--Infeld theory with spontaneously broken
$\mathcal N=4$ supersymmetry, verified its self-duality to tenth order,
and proposed a recursive construction of the action~\cite{IZ2014}.
Supersymmetric self-duality was formulated
in~\cite{KT2000,KTReview2001,KM2003}. Auxiliary-field formulations
express self-duality as phase invariance of the auxiliary interaction.
This approach was developed for bosonic electrodynamics
in~\cite{IZ2002,IZ2004,IZ2013} and extended to superspace by
Kuzenko~\cite{K2013} and, independently for $\mathcal N=1$, by Ivanov,
Lechtenfeld and Zupnik~\cite{ILZ2013}. Bandos, Lechner, Sorokin and
Townsend obtained the off-shell $\mathcal N=1$ ModMax action and its
super-Weyl invariant coupling~\cite{BLST2021}. Further superconformal
and nonlinear models appear in~\cite{K2021,KR2021,KS2023,KMcA2023},
including higher-derivative deformations~\cite{KR2024}.

The $\mathcal N=2$ vector multiplet decomposes into an $\mathcal N=1$
vector and a chiral multiplet $\Phi$. Setting $\Phi=0$ removes the
complex scalar, second gaugino and two real auxiliary fields. Agreement
with $\mathcal N=1$ ModMax therefore leaves their dynamics untested.
Our equivalence uses a real vector-prepotential redefinition after this
restriction; related nonlinear supersymmetric changes of variables were
studied in~\cite{KM2005,KTy2011}.

Kuzenko and Raptakis observed that an $\mathcal N=2$ superconformal
extension of ModMax had not been constructed, identifying the
dimensionless full $\mathcal N=2$ superspace measure as the main
difficulty, and presented a model involving a compensating vector
multiplet~\cite[section~4.3]{KR2021}. We compare the two constructions
in Section~\ref{sec:compensators}. In this paper we construct a family
of superconformal primary interactions built from the Abelian reduced
chiral field strength alone, solve the self-duality equation to all
orders with prescribed phase-invariant terms, and show that, with
$\Phi$ switched off, the resulting action is equivalent off shell to
the $\mathcal N=1$ ModMax action. We work in ordinary flat
$\mathcal N=2$ superspace and use the rigid superconformal Killing
formalism of~\cite{K2008}. No compensator or other additional multiplet
enters the action. The change of variables used to construct the completion
is the $\mathcal N=2$ auxiliary-superfield representation
of~\cite[section~4]{K2013}, also used in~\cite{IZ2014}, which we
rederive in our normalization.
A recent account of nonlinear self-duality for different amounts of supersymmetry is given in~\cite{K2026}.

\vspace{0.5cm}
In this paper, we construct a compensator-free $\mathcal N=2$
superconformal and self-dual extension of ModMax electrodynamics.
The action takes the form of a formal series:
\begin{equation}
 \boxed{\begin{gathered}
 \widetilde S_t=S_0+\sum_{n\geq1}t^n\widetilde F_n,\qquad
 t=\sinh\gamma,\qquad \widetilde F_1=f_{\rm mod},\\
 \Pi_0\widetilde F_n=[t^n]d(t)\,\Delta\quad(n\geq2),\\
 d(t)=\frac1{19}\left(\frac{2t}{\sqrt{1+t^2}+1}-t\right).
 \end{gathered}}
 \label{eq:proposal}
\end{equation}
Here $f_{\rm mod}$ is the first-order interaction of
Section~\ref{sec:additionalinteractions}, built from the minimal square-root
interaction $T_1$ and one further primary density. The combination
$\Delta=f_{\rm mod}-T_1$ is given in \eqref{eq:deltastar} and vanishes under
the $\mathcal N=1$ vector restriction $\Phi=0$ and therefore does not
alter the reduction to the Bandos action. The phase average $\Pi_0$ is defined in
\eqref{ao:eq:canonical}. The charged parts of the higher coefficients
follow from the self-duality equation: the displayed phase-invariant
series is not simply added to an already completed action.
For comparison we also consider the canonical completion, in which
all higher-order phase-invariant terms vanish. The choice of $d(t)$
comes from the zero-vev scalar analysis. Section~\ref{sec:rootflow}
shows that, within the quadratic scalar form used here and with
Maxwell initial data, the root-flow condition modulo equations of
motion fixes this function. These data do not uniquely select the
full superspace completion.

Two inductions establish the formal self-dual $\mathcal N=2$ action
and its off-shell $\mathcal N=1$ equivalence. The auxiliary-superfield
formalism is that of~\cite{K2013,IZ2014}; its use with prescribed phase averages,
and the all-order prepotential matching, are the steps needed here.

We study fluctuations about constant non-null electromagnetic backgrounds
at vanishing scalar expectation value. The chiral-multiplet fluctuations
are retained, so this calculation is distinct from setting $\Phi=0$. The minimal interaction
gives a superluminal scalar already at first order. One additional
primary density reverses this coupling without changing the
$\mathcal N=1$ reduction. For \eqref{eq:proposal}, the zero-vev scalar quadratic action
is determined to all orders and has an analytic continuation with
$v_\perp^2=e^{-\gamma}$ and $v_\parallel^2=1$ for finite $\gamma\geq0$.
On the same backgrounds, the $\mathcal N=1$ equivalence and $SU(2)_R$
symmetry determine both gaugino operators and the auxiliary triplet.
The gauginos remain luminal with positive kinetic coefficients, and
the auxiliary fields remain algebraic.

The characteristic analysis of nonlinear theories distinguishes a
second-order effective metric~\cite{BMV2008,ACGR2015,Reall2021} from
finite-order dispersive corrections and front velocities
\cite{Shore2007,CGdRPT2022}. Photon causality in self-dual electrodynamics
\cite{RT2023,RT2024a,RT2024b} does not settle the additional scalar
sector. Here the scalar quadratic operator is exactly second order
in derivatives on the specified backgrounds.

Stress-tensor flows offer another route to ModMax-type theories
\cite{BBYM2022,FST2022,FST2023}; marginal square-root deformations were also
studied in two dimensions~\cite{FSST2022}. The relation between
duality-preserving deformations and stress-tensor flows was established
in~\cite{FKST2023} for Lagrangians depending on the field strength but
not on its derivatives, and combined irrelevant and marginal flows
were studied in~\cite{BHO2024}. The photon and zero-vev scalar quadratic
sectors agree with the bosonic square-root flow, the latter modulo
equations of motion and total derivatives. This agreement selects the
scalar coefficient function; it does not independently establish a flow
for the full action. Supersymmetric flows involve complete conserved
supercurrent multiplets and their allowed improvements
\cite{BSTW2019,CFSST2020,FST2023}. Constructing such a flow is a separate
problem from the self-dual completion studied here.

Non-Abelian ModMax models with self-dual instantons, gravitating dyons
\cite{CCDGO2026} and merons~\cite{DG2026} have also been studied.
Instanton self-duality is distinct from the continuous electric--magnetic
duality considered here for an Abelian $\mathcal N=2$ vector multiplet.

The all-order construction is formal. Convergence of the full nonlinear
series and propagation on more general backgrounds remain open. The
finite-coupling statements refer to the analytic continuation of the
quadratic series at vanishing scalar expectation value.

Sections 2--5 give the formal construction; Sections 6--7 establish
the $\mathcal N=1$ equivalence. Section 8 contains the propagation
results and comparisons with earlier work. The appendices supply the
finite field-redefinition identities, the scalar recursion and the
superconformal variations.
The first-order interaction is specified in Section~\ref{sec:fmod};
Appendix~\ref{app:svariation} gives the superconformal variations.

\section{Conventions and constraints}
\label{sec:conventions}
\subsection{Flat superspace and the reduced-chiral constraint}
We work in ordinary four-dimensional $\mathcal N=2$ superspace with
coordinates $(x^\mu,\theta_i^\alpha,\bar\theta^{\dot\alpha i})$,
where $i=1,2$ is an $SU(2)$ internal isospin index. The flat spinor derivatives obey
\begin{equation}
 \{D^i_\alpha,\bar D_{\dot\beta j}\}
 =-2i\delta^i{}_j\partial_{\alpha\dot\beta},\qquad
 \{D^i_\alpha,D^j_\beta\}
 =\{\bar D_{\dot\alpha i},\bar D_{\dot\beta j}\}=0.
 \label{eq:flatalgebra}
\end{equation}
Internal indices are raised and lowered with the antisymmetric
$SU(2)$ tensors, with $\varepsilon_{12}=1$ and
$\varepsilon^{ij}\varepsilon_{jk}=\delta^i{}_k$. All spinor squares are normalized with a factor $\tfrac12$:
\begin{equation}
 (D^i)^2:=\tfrac12D^{\alpha i}D^i_\alpha\quad ,
 \qquad
 (\bar D_i)^2:=\tfrac12\bar D_{\dot\alpha i}\bar D_i^{\dot\alpha}.
 \label{eq:halfsquares}
\end{equation}
Thus $(D^2)^2$ denotes the square of the derivative with internal index
$i=2$; not to be confused with the $\mathcal N=1$ operator
$D^2=\tfrac12D^\alpha D_\alpha$ introduced only after the truncation.

Define the symmetric triplet operators in the same normalization by
\begin{equation}
 D^{ij}:=\tfrac12D^{\alpha(i}D^{j)}_\alpha,\qquad
 \bar D^{ij}:=\tfrac12\bar D_{\dot\alpha}^{(i}
                           \bar D^{\dot\alpha j)}.
 \label{eq:tripletoperators}
\end{equation}
Parentheses indicate symmetrization with unit weight. The Abelian
vector-multiplet field strength $\W$ is an $SU(2)_R$ singlet satisfying
\begin{equation}
 \bar D_{\dot\alpha i}\W=0,\qquad
 D^{ij}\W=\bar D^{ij}\bW.
 \label{eq:reducedchiral}
\end{equation}
The second relation is the Bianchi identity defining the reduced-chiral
field strength and holds independently of the equations of motion.
For the free Maxwell theory, the equation of motion additionally implies
$D^{ij}\W=0$. The reduced-chiral description is reviewed in~\cite{BK2010}. The derivative with $i=1$ becomes the manifest
$\mathcal N=1$ derivative only when the truncation to the $\mathcal N=1$ vector multiplet is made in
Section~\ref{sec:n1}.

\subsection{Measures and functional differentiation}
We write the spacetime and Grassmann integration measures explicitly. To abbreviate
repeated functional pairings, define
\begin{equation}
 \begin{aligned}
 \IC[A]&=\int\dd^4x\,\dd^4\theta\,A,\qquad
 \IF[B]=\int\dd^4x\,\dd^8\theta\,B,\\
 \IF[B]&=\IC[\overline{\mathscr P}B],\qquad
 \overline{\mathscr P}:=(\bar D_1)^2(\bar D_2)^2.
 \end{aligned}
 \label{eq:measures}
\end{equation}
Here $\dd^4\theta$ is the chiral $\mathcal N=2$ measure and
$\dd^8\theta$ is the full $\mathcal N=2$ measure. The last equality
fixes their relative normalization. The product of barred squares is invariant under constant
$SU(2)_R$ changes of basis and maps an unconstrained scalar to a chiral
one. The $\mathcal N=1$ measure normalization used after the truncation
is in \eqref{sr:eq:projection}.

The free action and the definition of a chiral functional derivative are
\begin{equation}
 S_0=\frac18\big(\IC[\W^2]+\cc\big),\qquad
 \delta F=\IC\left[\delta\W\,\frac{\delta F}{\delta\W}\right]
       +\overline{\IC}\left[\delta\bW\,\frac{\delta F}{\delta\bW}\right].
 \label{eq:maxwellgradient}
\end{equation} To define the functional derivatives, we treat $\W$ and $\bW$ as
independent chiral and antichiral superfields, without imposing the
Bianchi identity~\cite{KT2000}. We differentiate the explicit superspace
expressions given below. The Bianchi identity is imposed only afterwards, on physical configurations.

\subsection{Notation}
\label{sec:notation}
Several letters have different meanings in separate parts of the paper.
Each is defined where it is introduced; the recurring cases are listed
below.
\begin{center}
\small
\begin{tabular}{@{}p{0.2\textwidth}p{0.76\textwidth}@{}}
\toprule
Symbol & Meaning (where used)\\
\midrule
$\PP(F,G)$, $\overline{\mathscr P}$ &
 pairing $\IC[F_{\W}G_{\W}]$ of chiral functional derivatives
 (Section~\ref{sec:recursion}); chiral projector
 \eqref{eq:measures}\\
$P$ &
 free wave operator $\omega^2-\mathbf q^2$
 (Appendix~\ref{app:components})\\
$\HH$, $H_n$ &
 auxiliary functional and its phase-invariant coefficients
 (Sections~\ref{sec:recursion} and \ref{sec:additionalinteractions})\\
$K_n$, $\hK$, $K_{pq}$ &
 $K_n=[t^n]I^{\rm trial}$ (Section~\ref{sec:induction});
 primary tensors $\hK$ and $K_{pq}$
 (Sections~\ref{sec:primary} and \ref{sec:additionalinteractions})\\
$G$ &
 truncated density $\mathcal G(t;c,y)$ (Section~\ref{sec:matching});
 descendants $G^i,G^{ij}$ (Section~\ref{sec:primary} and
 Appendix~\ref{app:svariation})\\
$f$, $g$ &
 background field-strength amplitudes \eqref{eq:componentdictionary};
 in Section~\ref{sec:recursion}, a general first-order interaction and
 $g=I_{\W}$\\
$\nu$, $\varpi$ &
 $\nu=(f^2+g^2)/(2fg)=-x/r$ \eqref{eq:componentdictionary};
 $\varpi=\sqrt{1+2cy+y^2}$ (Section~\ref{sec:matching})\\
$u$ &
 composite of \eqref{sr:eq:composites}; coefficient function
 $u(\vartheta,t)$ (Section~\ref{sec:zeromultiplet})\\
$p$ &
 electromagnetic invariant \eqref{eq:bosmodmax}; polarized derivative
 $p_y$ (Section~\ref{sec:primary}); momentum matrix
 (Appendix~\ref{app:components})\\
$T$ &
 bilinear $T_{vw}$ (Section~\ref{sec:primary}); coefficients $T_n$
 (Section~\ref{sec:recursion}); Maxwell stress tensor $T_F$\\
\bottomrule
\end{tabular}
\end{center}

\section{Superconformal invariance of the minimal interaction}
\label{sec:primary}
\subsection{The bilinear ansatz}
Let $v=(\alpha\dot\alpha)$ and $w=(\beta\dot\beta)$. The operator $\Pi$
symmetrizes undotted and dotted index pairs separately, each with unit
weight. Define
\begin{equation}
 \begin{aligned}
 X_{vw}&=\Pi\big[(\partial_v\partial_w\W)\bW
                          +\W\partial_v\partial_w\bW\big],\\
 Y_{vw}&=\Pi(\partial_v\W)(\partial_w\bW),\\
 Z_{vw}&=\Pi\sum_i\big[(\partial_vD^i_\beta\W)(\bar D_{\dot\beta i}\bW)
               -(D^i_\beta\W)\partial_v(\bar D_{\dot\beta i}\bW)\big],\\
 T_{vw}&=\Pi\sum_{i,j}(D^i_\alpha D^j_\beta\W)
                            (\bar D_{\dot\alpha i}\bar D_{\dot\beta j}\bW).
 \end{aligned}
 \label{eq:basis}
\end{equation}The symbol $T_{vw}$ in this basis is not the physical stress tensor. All four
structures have dimension four and zero $U(1)$ charge; $X,Y,T$ and $iZ$ are
real under graded conjugation with the displayed spinor-index ordering. The
same bilinear sector occurs in the higher-spin currents of~\cite{KR2023}; see
Section~\ref{sec:currentscomparison}. Here the coefficients are fixed by
superconformal invariance off shell, without using a conservation law or
identifying the tensor with the stress tensor obtained by metric variation.

For the superconformal variation, contract the tensor indices with commuting spinors
$y^\alpha,\bar y^{\dot\alpha}$ and use
\begin{equation}
 \begin{gathered}
 p_y=y^\alpha\bar y^{\dot\alpha}\partial_{\alpha\dot\alpha},\qquad
 D^i(y)=y^\alpha D^i_\alpha,\quad
 \bar D_i(\bar y)=\bar y^{\dot\alpha}\bar D_{\dot\alpha i},\\
 F^i=D^i(y)\W,\qquad G_i=\bar D_i(\bar y)\bW.
 \end{gathered}
 \label{eq:polarization}
\end{equation}
In the following polarized calculation we write $D^i(y)$ and
$\bar D_i(\bar y)$ as $D^i$ and $\bar D_i$. Products such as
$D^iD^j$ contain two polarized first-order derivatives, not the
contracted square of \eqref{eq:halfsquares}. The polarization
spinors are unrelated to the real variable $y$ used later in
the truncation to the $\mathcal N=1$ vector multiplet.

\subsection{Constrained superconformal parameters}
The real superconformal Killing vector $\xi$ preserves chirality
and satisfies~\cite{K2008}
\begin{equation}
 \begin{gathered}
 [\xi,D^i_\alpha]=\omega_\alpha{}^\beta D^i_\beta
                    -\bar\sigma D^i_\alpha-\Lambda_j{}^iD^j_\alpha,
 \qquad\bar D_{\dot\alpha i}\sigma
 =\bar D_{\dot\alpha i}\omega_\beta{}^\gamma=0,\\
 D^k_\alpha\Lambda_j{}^i
 =-2\left(\delta_j{}^kD^i_\alpha
                  -\tfrac12\delta_j{}^iD^k_\alpha\right)\sigma.
 \end{gathered}
 \label{eq:killing}
\end{equation}
Reality, the conjugate identities and the traceless $SU(2)$ condition
complete the constraints. These parameter functions are determined by
the finite set of rigid $SU(2,2|2)$ parameters; they are not arbitrary
local transformation parameters. The field strength transforms as
\begin{equation}
 \delta_{\rm sc}\W=-\xi\W-2\sigma\W,\qquad
 \delta_{\rm sc}\bW=-\xi\bW-2\bar\sigma\bW.
 \label{eq:Wprimary}
\end{equation}
It preserves the reduced-chiral representation.

For the measure conventions of \eqref{eq:measures}, constant
normalization factors do not alter the transformation laws of the measures:
\begin{equation}
 \mathcal L_\xi(\dd^4x\,\dd^8\theta)=0,\qquad
 \mathcal L_\xi(\dd^4x\,\dd^4\theta)
       =4\sigma\,\dd^4x\,\dd^4\theta.
 \label{eq:measurevariation}
\end{equation}The full superspace measure is invariant because the contributions of the four
even and eight odd coordinates to the superdivergence of $\xi$ cancel. The
transformation of the chiral measure makes \eqref{eq:maxwellgradient} invariant. Integration by
parts is assumed to have no spacetime boundary term.

The $SU(2)_R$ transformations act on the internal derivative indices.
Every internal index in the four bilinears is contracted, so the tensor
is an $SU(2)_R$ singlet. The remaining test is the cancellation of the
inhomogeneous terms produced when derivatives act on the constrained
superconformal parameters.

\subsection{Special-supersymmetry calculation}
We denote by $\AS$ the inhomogeneous part of a special-supersymmetry variation: the remainder after subtracting the transport term, the Lorentz and $R$-symmetry rotations, and the homogeneous scaling terms. For pure special
supersymmetry put
\begin{equation}
 \chi^i=D^i\sigma,\qquad\bar\chi_i=\bar D_i\bar\sigma.
 \label{eq:oddparameters}
\end{equation}
They are constant odd parameters in the polarized calculation.
The Killing identities give $D^i\omega_y{}^\beta=-\chi^i y^\beta$;
the derivatives of the $SU(2)_R$ parameter are fixed by
\eqref{eq:killing}. Differentiating \eqref{eq:Wprimary}, including
these parameter derivatives, gives
\begin{equation}
 \begin{aligned}
 \AS(F^i)&=-2\chi^i\W,&
 \AS(G_i)&=-2\bar\chi_i\bW,\\
 \AS(p_y\W)&=-i\bar\chi_iF^i,&
 \AS(p_y\bW)&=-i\chi^iG_i,\\
 \AS(p_yF^i)&=-4\chi^ip_y\W-i\bar\chi_jD^jD^i\W,&
 \AS(p_yG_i)&=-4\bar\chi_ip_y\bW-i\chi^j\bar D_j\bar D_i\bW,\\
 \AS(D^iD^j\W)&=4(\chi^jF^i-\chi^iF^j),&
 \AS(\bar D_i\bar D_j\bW)&=4(\bar\chi_jG_i-\bar\chi_iG_j).
 \end{aligned}
 \label{eq:jetvariations}
\end{equation}
The remaining variations organize into three structures:
\begin{equation}
 \begin{aligned}
 U_S&=\chi^i\W p_yG_i+\bar\chi_i(p_yF^i)\bW,\\
 V_S&=\chi^i(p_y\W)G_i+\bar\chi_iF^ip_y\bW,\\
 M_S&=\chi^iF^j(\bar D_i\bar D_j\bW)+\bar\chi_i(D^iD^j\W)G_j.
 \end{aligned}
 \label{eq:anomalybasis}
\end{equation}
Applying the graded product rule gives
\begin{equation}
 \AS(X)=-2iU_S,\quad \AS(Y)=-iV_S,\quad
 \AS(Z)=2U_S-4V_S-iM_S,\quad \AS(T)=-8M_S.
 \label{eq:basisvariation}
\end{equation}
Appendix~\ref{app:svariation} derives the terms containing derivatives
of the transformation parameters in \eqref{eq:jetvariations} and gives the product expansions
leading to every entry of \eqref{eq:basisvariation}. 
\subsection{The primary tensor and the variation of the action}
For $\hK=c_XX+c_YY+c_ZZ+c_TT$, cancellation of
\eqref{eq:basisvariation} requires
\begin{equation}
 -2ic_X+2c_Z=0,\qquad -ic_Y-4c_Z=0,\qquad -ic_Z-8c_T=0.
 \label{eq:coeffconstraints}
\end{equation}
Requiring the tensor to be primary fixes the solution up to scale
within this four-term basis.
Normalizing $c_T=-4$ gives
\begin{equation}
 \boxed{\hK=-32X+128Y-32iZ-4T.}
 \label{eq:primarytensor}
\end{equation}The sum of its inhomogeneous terms is
$64iU_S-128iV_S-64iU_S+128iV_S-32M_S+32M_S=0$. Translations, $Q$ supersymmetry
and constant Lorentz and $R$ transformations are already respected by
construction. Invariance under $S$ and $\bar S$ at the origin, together with
the superconformal algebra, implies invariance under special conformal
transformations. Thus the full rigid law is\begin{equation}
 \delta_{\rm sc}\hK_{vw}=-\xi\hK_{vw}
 +L_v{}^u\hK_{uw}+L_w{}^u\hK_{vu}
 -4(\sigma+\bar\sigma)\hK_{vw},
 \label{eq:fullKlaw}
\end{equation}
where $L_{\alpha\dot\alpha}{}^{\beta\dot\beta}
=\omega_\alpha{}^\beta\delta_{\dot\alpha}{}^{\dot\beta}
+\delta_\alpha{}^\beta\bar\omega_{\dot\alpha}{}^{\dot\beta}$.

On the domain where the body of $\hK\cdot\hK$ is nonzero, and with a fixed branch of the square root, define
\begin{equation}
 Q=\sqrt{\hK\cdot\hK},\qquad
 \boxed{T_1=\kappa
 \int\dd^4x\,\dd^8\theta\,\frac{\W^2\bW^2}{Q}.}
 \label{eq:minimalinteraction}
\end{equation}

Here $\hK\cdot\hK=\hK_{vw}\hK^{vw}$ denotes contraction of the
tensor indices. The coefficient $\kappa$ is fixed by requiring the
photon sector to reproduce the ModMax Lagrangian with the positive
branch of the square root. On the constant electromagnetic backgrounds
of Appendix~\ref{app:components}, with field-strength parameters
$f,g$ defined in \eqref{eq:componentdictionary}, we obtain
\begin{equation}
 Q_0=16fg,\qquad
 [\W^2\bW^2]_{\theta^4\bar\theta^4}=4f^2g^2,\qquad
 \boxed{\kappa=2}.
 \label{eq:calibration}
\end{equation}
Thus $T_1$ has constant-field density $fg/2$ in these conventions;
$\kappa$ is not an independent coupling.

Inverse powers of a superfield require its lowest commuting component
to be nonzero; square roots additionally require a choice of branch.
Accordingly, the interaction is defined where
$\operatorname{body}(\hK\cdot\hK)\ne0$. On the physical backgrounds
considered here,
\begin{equation}
 r=\sqrt{x^2+p^2}>0,\qquad
 \operatorname{body}(\hK\cdot\hK)=(16fg)^2>0,
 \label{eq:nonnull}
\end{equation}
where $g=f^*$. We choose the positive square root on these backgrounds.
For varying physical configurations, we restrict attention to regions
where this branch remains real and the denominators remain nonzero.
For independent complex chiral and antichiral arguments, we use its
local analytic continuation. Any additional denominators introduced
below must likewise remain nonzero.

The scalar expectation value may vanish. However, constant
configurations with $F_{\mu\nu}=0$ and vanishing fermions and auxiliaries
have $\hK=0$, irrespective of the constant scalar value, and are
excluded. The coupling expansion is therefore defined on the domain
specified above, rather than as a weak-field expansion about the
zero-field vacuum.

Numerator and denominator have
the same homogeneous variation. Consequently
\begin{equation}
 \delta_{\rm sc}\left(\frac{\W^2\bW^2}{Q}\right)
 =-\xi\left(\frac{\W^2\bW^2}{Q}\right),\qquad
 \delta_{\rm sc}T_1=0.
 \label{eq:rootinvariance}
\end{equation}More generally, let $S_K$ denote the functional \eqref{eq:minimalinteraction}
with $\hK$ replaced by a denominator tensor $K_{vw}$ of the same weights.
If $K_{vw}$ has inhomogeneous variation $A_{vw}$, then\begin{equation}
 \delta S_K
 =-\kappa\IF\left[
 \frac{\W^2\bW^2}{(K\cdot K)^{3/2}}K^{vw}A_{vw}\right].
 \label{eq:wardfunctional}
\end{equation}
For \eqref{eq:primarytensor}, $A=0$ off shell.

\subsection{Chiral functional derivative and phase invariance}
Let $\mathscr L_{\bW}\chi=\hK[\chi,\bW]$ be the linearization in the
first argument. Define its graded formal adjoint by full superspace
integration by parts. The chiral functional derivative of $T_1$ is the local expression
\begin{equation}
 \boxed{s:=\frac{\delta T_1}{\delta\W}
 =\kappa(\bar D_1)^2(\bar D_2)^2\left[
 \frac{2\W\bW^2}{Q}
 -\mathscr L_{\bW}^{\dagger}
       \left(\frac{\W^2\bW^2\hK}{Q^3}\right)\right].}
 \label{eq:firstvariation}
\end{equation}
The adjoint is a finite differential operator determined by
\eqref{eq:basis}; it is not an inverse operator. Each recursive coefficient
below is therefore a specified local functional on the same regular domain.

Regarded as a functional of unconstrained chiral superfields, $T_1$ is
invariant under the rigid $U(1)$ phase rotation:\begin{equation}
 T_1[e^{i\alpha}\W,e^{-i\alpha}\bW]=T_1[\W,\bW],\qquad
 \im\IC[\W s]=0.
 \label{eq:firstphase}
\end{equation}This is the first-order duality condition. The higher-order conditions
determine the completion constructed in the next section.

\section{Canonical self-dual completion to all orders}
\label{sec:recursion}
\subsection{The functional equation and the normalization}
The $\mathcal N=2$ self-duality equation~\cite{KT2000,KTReview2001} is
\begin{equation}
 i\M=4\frac{\delta S}{\delta\W},\qquad
 \mathcal D[S]:=\im\IC[\W^2+\M^2]=0.
 \label{eq:GZ}
\end{equation}
It expresses covariance of the equations of motion and Bianchi identities
under $\delta\W=\lambda\M$, $\delta\M=-\lambda\W$. It is not the
off-shell invariance of $S$ under a linear phase rotation of $\W$.
The two-term expression $S_{\rm hyp}=\cosh\gamma\,S_0+
\sinh\gamma\,T_1$ does not satisfy this equation. With
$\mathscr S=4(T_1)_{\W}$, phase invariance of $T_1$ gives
\begin{equation}
 \mathcal D[S_{\rm hyp}]
 =-\sinh^2\gamma\,\im\IC[\W^2+\mathscr S^2].
 \label{eq:hyperboliccriterion}
\end{equation}
To exhibit the obstruction, take constant chiral fields with only
quadratic Grassmann components, $\W=\W_f+\W_y$, in the Lorentz
and auxiliary-triplet representations. Their chiral pairing is
orthogonal. Per unit spacetime volume put
$u_0=\langle\W_f^2\rangle_C$ and $v_0=\langle\W_y^2\rangle_C$.
The denominator depends only on $u_0\bar u_0$, and the photon
normalization \eqref{eq:calibration} fixes
\begin{equation}
 L_1=\frac{|u_0+v_0|^2}{4\sqrt{u_0\bar u_0}},\qquad
 \mathscr S=8\bigl(L_{1,u_0}\W_f+L_{1,v_0}\W_y\bigr).
 \label{eq:hyperbolicsector}
\end{equation}
Imposing the auxiliary-field reality condition $v_0=\bar v_0$ gives
$\im\langle\W^2+\mathscr S^2\rangle_C
=\im u_0[1-|u_0+v_0|^4/|u_0|^4]$.
For $u_0=i\rho$, $v_0=\epsilon\rho$, with $\rho>0$ and real
$\epsilon\ne0$ with the sign required by that reality condition,
\begin{equation}
 \mathcal D[S_{\rm hyp}]/\mathrm{Vol}
   =\sinh^2\gamma\,\rho(2\epsilon^2+\epsilon^4)\ne0.
 \label{eq:explicitobstruction}
\end{equation}
The failure already at $\gamma^2$ motivates the recursive completion.

For $S=S_0+I$ and $g=I_{\W}$ the equation reads
\begin{equation}
 \W^2+\M^2=-8\W g-16g^2,\qquad
 \delta_0 I=4\im\IC[g^2],\qquad
 \delta_0I=-2\im\IC[\W I_{\W}].
 \label{eq:interactionduality}
\end{equation}The phase rotation acts on unconstrained chiral superfields, and the same
extension is used in all subsequent pairings. Different extensions give
different results for functionals that depend on divergences of the field
strength~\cite{KS2023}.

Three $U(1)$ actions must be distinguished. Electric--magnetic duality
acts on the equations of motion and Bianchi identities through
$\delta\W=\lambda\M$, $\delta\M=-\lambda\W$. The rigid phase
rotation of unconstrained chiral superfields,
$\W\mapsto e^{i\alpha}\W$, $\bW\mapsto e^{-i\alpha}\bW$, defines
the phase-invariant functionals and the phase average $\Pi_0$ used below.
For the free action $i\M=\W$, so this phase rotation agrees with a
duality rotation with parameter $\lambda=-\alpha$; they differ in
the interacting theory. Finally, the $R$-symmetry
$U(1)_R\subset SU(2,2|2)$ also rotates the Grassmann coordinates and
is used in Section~\ref{sec:scalarpropagation} to exclude scalar--photon
mixing. In what follows, phase invariance refers to the rigid phase
rotation of the chiral arguments, and duality invariance refers to
electric--magnetic duality.

Let $\delta_0\W=i\W$ and $\delta_0\bW=-i\bW$. Define the phase average
\begin{equation}
 \Pi_0F:=\frac1{2\pi}\int_0^{2\pi}\dd\alpha\,
 F[e^{i\alpha}\W,e^{-i\alpha}\bW],
 \label{ao:eq:canonical}
\end{equation}
which projects $F$ onto its phase-invariant (charge-zero) part.
For the canonical completion we impose the normalization
\[
 T_1=f,\qquad\Pi_0f=f,\qquad\Pi_0T_n=0\quad(n\geq2).
\]
In this section $f$ denotes an arbitrary first-order interaction, with $T_1=f$ and
$I_t=\sum_{n\geq1}t^nT_n$. The canonical completion of $T_1$ is obtained by
choosing $f$ to be \eqref{eq:minimalinteraction}; the modified family is specified
in Section~\ref{sec:additionalinteractions}. Equalities of functionals are understood modulo superspace integration
by parts. Each coefficient is local; the number of derivatives need not
be uniformly bounded as $n$ grows. The power series is formal in $t$,
not an expansion about a vanishing field strength. The nonanalytic
field dependence of the first-order interaction is permitted on its chosen regular domain.

Put $g_n=(T_n)_{\W}$ and $\PP(F,G)=\IC[F_{\W}G_{\W}]$.
Substitution of the series in \eqref{eq:interactionduality} gives
\begin{equation}
 \begin{gathered}
 i\M=4(S_t)_{\W}=\W+4(I_t)_{\W},\qquad
 \mathcal D[S_t]:=\im\IC[\W^2+\M^2]=0,\\
 \delta_0T_n=4\im R_n,\qquad
 R_n=\sum_{k=1}^{n-1}\PP(T_k,T_{n-k}),\qquad
 \im\Pi_0R_n=0.
 \end{gathered}
 \label{ao:eq:recursion}
\end{equation}The last equality is an integrability condition: it must hold before
$\delta_0$ can be inverted on the $U(1)$-charged part.

The main result of this section is that the self-duality equation can be
solved order by order with this normalization. For every real, local,
phase-invariant first-order interaction $f$ on a regular domain preserved
by the phase rotation, there is a unique formal interaction $I_t$
satisfying this normalization and \eqref{ao:eq:recursion}. If $f$ is
invariant under rigid $SU(2,2|2)$ transformations of unconstrained chiral
superfields, so is every $T_n$. We establish existence and uniqueness in
Sections~\ref{sec:auxiliary} and \ref{sec:induction}, and superconformal
invariance in Section~\ref{sec:covariance}.

\subsection{Auxiliary-superfield form of the self-duality equation}
\label{sec:auxiliary}
For any real $I=O(t)$, introduce the auxiliary chiral superfield and
the transformed functional
\begin{equation}
 \UU=\W+2g,\qquad g=I_{\W},\qquad
 \HH[\UU,\bar\UU]
 =I[\W,\bW]+\IC[g^2]+\cc
 \label{ao:eq:transform}
\end{equation}Here only the square term is conjugated in the last expression. The inverse
$\W=\W[\UU,\bar\UU]$ is substituted on its right-hand side. It exists formally
because $\UU=\W+O(t)$. The composite $\UU$ is an auxiliary chiral variable,
with no reduced-chiral constraint. The transformation acts on unconstrained
chiral superfields and leaves the physical vector multiplet unchanged.

Vary \eqref{ao:eq:transform}, keeping all functional integrations and
adjoints. Since these chiral functional derivatives are Grassmann even,
\begin{equation}
 \begin{aligned}
 \delta\HH
 &=\IC[g\,\delta\W+2g\,\delta g]+\cc
  =\IC[g\,\delta\UU]+\cc,
 &\HH_{\UU}&=g,\\
 \delta_0^{\UU}\HH
 &=-2\im\IC[\UU\HH_{\UU}]
  =-2\im\IC[\W g+2g^2]
  =\tfrac14\mathcal D[S_0+I].
 \end{aligned}
 \label{ao:eq:linearization}
\end{equation}The phase operator on the second line acts linearly on $\UU$, not on $\W$ at
fixed $I$. The final equality follows directly from $\W^2+\M^2=-8\W g-16g^2$.
Thus self-duality is equivalent to ordinary phase invariance of $\HH$. This is
the $\mathcal N=2$ auxiliary-superfield formulation of~\cite{K2013,IZ2014}
in our normalization. The phase-invariance condition is given explicitly
in Eqs.~(3.11)--(3.14) of~\cite{IZ2014}. The corresponding bosonic
construction uses the auxiliary-bispinor
formulation of~\cite{IZ2002,IZ2004,IZ2013}. Here $\UU$ is determined
by \eqref{ao:eq:transform}; the physical action depends on $\W$ and
$\bW$ alone.

Conversely, prescribe a real phase-invariant $\HH=O(t)$ and solve
\begin{equation}
 \UU=\W+2\HH_{\UU}[\UU,\bar\UU],\qquad
 I[\W,\bW]=\HH[\UU,\bar\UU]
              -\big(\IC[(\HH_{\UU})^2]+\cc\big).
 \label{ao:eq:inverse}
\end{equation}
One determines the $t^n$ coefficient of $\UU$ from the previous
coefficients because $\HH_{\UU}=O(t)$. No inverse spacetime or
spinor differential operator is used. Differentiating the second
formula yields $I_{\W}=\HH_{\UU}$: its variation is
$\IC[\HH_{\UU}(\delta\UU-2\delta\HH_{\UU})]+\cc$
and the bracket equals $\delta\W$. Therefore the two transformations
are inverse to each other and the generated $I$ is self-dual.

For the canonical normalization we need a further property. At fixed
lower coefficients, adding a functional $J$ at order $n$ gives
\begin{equation}
 \HH\longmapsto\HH+t^nJ
 \quad\Longrightarrow\quad
 I\longmapsto I+t^nJ[\W,\bW]+O(t^{n+1}).
 \label{ao:eq:triangular}
\end{equation}
Indeed, the change of $\UU$ starts at order $n$. Its contribution to
$\HH$ is multiplied by $\HH_{\UU}=O(t)$, and every cross term in
the square of the functional derivative also starts at order $n+1$. The direct contribution
is $t^nJ(\UU)=t^nJ(\W)+O(t^{n+1})$. Thus the normalization can be imposed order by order without changing
the lower-order action.

\subsection{Inductive construction and uniqueness}
\label{sec:induction}
Start with $H_1=f$ in $\HH=\sum_{j\geq1}t^jH_j$. The action obtained
from \eqref{ao:eq:inverse} has $T_1=f$. Suppose real phase-invariant
$H_1,\ldots,H_{n-1}$ have been chosen so the action has the required
canonical coefficients below order $n$. Form the finite polynomial
$\HH^{<n}=\sum_{j=1}^{n-1}t^jH_j$ and let $I^{\rm trial}$ be its
formal interaction obtained from \eqref{ao:eq:inverse}. Define
\begin{equation}
 K_n=[t^n]I^{\rm trial},\qquad
 H_n=-\Pi_0K_n,\qquad
 \boxed{T_n=K_n-\Pi_0K_n.}
 \label{ao:eq:induction}
\end{equation}
In $H_n$ the arguments of the functional $-\Pi_0K_n$ are renamed
$\UU,\bar\UU$; the physical coefficient $T_n$ retains $\W,\bW$.
The polynomial $\HH^{<n}$ is phase invariant, so the entire trial
interaction is self-dual, including its order-$n$ coefficient $K_n$.
The new $H_n$ is real and phase-invariant. Equation~\eqref{ao:eq:triangular}
leaves all lower coefficients unchanged and gives precisely the boxed
$T_n$. It obeys $\Pi_0T_n=0$, completing the induction.

This also establishes the integrability condition without a
separate high-order Hessian expansion. At order $n$ the exact trial
action satisfies $\delta_0K_n=4\im R_n$, where $R_n$ is constructed
only from the already fixed $T_1,\ldots,T_{n-1}$. Therefore
\begin{equation}
 \boxed{0=\Pi_0\delta_0K_n=4\im\Pi_0R_n
 \qquad\text{for every }n\geq2.}
 \label{ao:eq:neutral}
\end{equation}
The trial action exists by the explicit inversion \eqref{ao:eq:inverse}, independently of
the phase recursion we are proving solvable.

Uniqueness follows in the same way. The lowest-order difference $\Delta T_n$ of
two canonical solutions obeys $\delta_0\Delta T_n=0$ because their
lower coefficients, and hence their source $R_n$, agree. Consequently
$\Delta T_n=\Pi_0\Delta T_n=0$. Thus the induction constructs the
same solution as phase decomposition,
$T_n=\sum_{m\ne0}(-2R_{n,m}/m+\cc)$ with
$\delta_0R_{n,m}=imR_{n,m}$.

\subsection{First coefficients and superconformal covariance}
\label{sec:covariance}
As a normalization check, let $s=f_{\W}$,
$B=\PP(f,f)=\IC[s^2]$ and $C=\PP(f,B)$. The first coefficients are
\begin{equation}
 \boxed{T_1=f,\qquad T_2=B+\bar B,\qquad T_3=C+\bar C.}
 \label{eq:loworders}
\end{equation}
Choosing only $\HH=tf$ in the inverse transformation instead gives
\begin{equation}
 I\big[\HH=tf\big]
 =tf+t^2(B+\bar B)
 +t^3\big[C+\bar C+2\PP(f,\bar B)\big]+O(t^4).
 \label{eq:trialcubic}
\end{equation}
Symmetry of the mixed functional Hessian of real $f$ makes
$\PP(f,\bar B)$ real; its $U(1)$ charge is zero. The induction therefore
selects $H_3=-2\PP(f,\bar B)$, with arguments renamed to $\UU$, and
removes exactly this term from the physical coefficient. A
phase-invariant auxiliary interaction generates a self-dual action,
while this additional subtraction selects the canonical completion.

If a functional $F$ is superconformally invariant on unconstrained chiral
superfields, its chiral functional derivative transforms in the same primary
representation as $\W$. Its first variation and the chiral measure then imply\begin{equation}
 \begin{gathered}
 \delta_{\rm sc}F_{\W}=-\xi F_{\W}-2\sigma F_{\W},\qquad
 \delta_{\rm sc}\PP(F,G)=0
 \quad(F,G\text{ invariant}),\\
 \boxed{\delta_{\rm sc}T_n=0\ (n\geq1),\qquad
 \delta_{\rm sc}S_t=0,\qquad\mathcal D[S_t]=0}
 \quad\text{as formal series.}
 \end{gathered}
 \label{ao:eq:allordersymmetry}
\end{equation}
Phase decomposition commutes with the superconformal action, so inverting $\delta_0$ on terms with nonzero phase charge preserves
invariance at each order. Equivalently,
\eqref{ao:eq:inverse} is covariant and \eqref{ao:eq:induction} preserves
invariance. Finite differentiation and contraction also preserve
locality at each order. The physical Bianchi identity is not imposed
on a functional derivative or on $\UU$ to obtain these results.

\section{Additional primary interactions and the phase-invariant terms}
\label{sec:additionalinteractions}
\subsection{Primary tensors with unequal weights}
The superconformal variation fixes the tensor within the four-term ansatz, but it does
not exhaust the superconformal scalar interactions. Let $A=\W^p$ and
$\bar B=\bW^q$. In the polarized notation of Section~\ref{sec:primary}, set
\begin{equation}
 \begin{gathered}
 X_L=(p_y^2A)\bar B,\quad X_R=A p_y^2\bar B,\quad
 Y=(p_yA)(p_y\bar B),\\
 Z_L=(p_yD^iA)(\bar D_i\bar B),\quad
 Z_R=(D^iA)(p_y\bar D_i\bar B),\quad
 T=(D^iD^jA)(\bar D_i\bar D_j\bar B).
 \end{gathered}
 \label{eq:mixedbasis}
\end{equation}
To restore the Lorentz indices, one symmetrizes the two undotted and the two dotted
Lorentz indices separately. The resulting primary tensor is
\begin{equation}
 K_{pq}=4q(q+1)X_L+4p(p+1)X_R-8(p+1)(q+1)Y
       +4i(q+1)Z_L-4i(p+1)Z_R+T.
 \label{eq:mixedprimary}
\end{equation}
It has dimension $p+q+2$ and $U(1)$ charge $p-q$, with $\hK=-4K_{11}$.
The six inhomogeneous special-supersymmetry coefficients, in the ordered
basis of \eqref{eq:mixedbasis}, are the rows of
\begin{equation}
 \begin{pmatrix}
 0&-2i&0&0&-2p&0\\
 -2i&0&0&2q&0&0\\
 0&0&-i&-2(p+1)&0&0\\
 0&0&-i&0&2(q+1)&0\\
 0&0&0&0&i&-4(p+1)\\
 0&0&0&-i&0&-4(q+1)
 \end{pmatrix}.
 \label{eq:mixedwardmatrix}
\end{equation}
They multiply the ordered structures
\[
\begin{gathered}
\chi^iA p_y\bar D_i\bar B,\quad
\bar\chi_i(p_yD^iA)\bar B,\quad
\chi^i(p_yA)\bar D_i\bar B,\\
\bar\chi_iD^iA p_y\bar B,\quad
\chi^iD^jA\bar D_i\bar D_j\bar B,\quad
\bar\chi_iD^iD^jA\bar D_j\bar B.
\end{gathered}
\]
Multiplying \eqref{eq:mixedwardmatrix} by the coefficient vector in
\eqref{eq:mixedprimary} gives zero in every row. These are the weighted
versions of the calculation of the parameter derivatives in
Appendix~\ref{app:svariation}. They use only the chirality of $\W^p$. Together with the homogeneous transformation law
and the superconformal algebra, this establishes primarity.

\subsection{The first-order interaction}
\label{sec:fmod}
For $p=q=3$ the tensor $K_{33}$ has dimension eight and zero $U(1)$
charge. Contracting it with $\hK$ gives a real, phase-invariant density
whose superspace integrand has dimension zero,
\begin{equation}
 D_{33}=\IF\!\left[\frac{\hK^a{}_b(K_{33})^b{}_a}{Q^3}\right],
 \label{eq:additionaldensities}
\end{equation}
where the contracted indices are Lorentz vector indices. The interaction
used in \eqref{eq:proposal} is
\begin{equation}
 \boxed{f_{\rm mod}=T_1-\frac{152}{405}\,D_{33},}
 \label{eq:additionalinteractions}
\end{equation}
with the normalization $T_1=2\IF[\W^2\bW^2/Q]$. Both terms are real, phase
invariant and superconformal. The interaction therefore satisfies the
first-order self-duality equation for independent chiral arguments.
The interaction $D_{33}$ has zero $\mathcal N=1$ vector restriction, as shown
in Section~\ref{sec:modifiedrestriction}.

At vanishing scalar expectation value, the scalar Hessians of $T_1$ and
$D_{33}$ are $-18\Sigma$ and $-\frac{405}{8}\Sigma$ in the conventions of
Appendix~\ref{app:components}, and all their other independent-chiral
scalar entries vanish. The coefficient $-152/405$ is fixed by requiring
the total coefficient $\Sigma$: the additional term contributes $19\Sigma$.
It changes the first-order scalar correction from superluminal to subluminal without
changing the $\mathcal N=1$ reduction.

The results of Sections~\ref{sec:matching} and \ref{sec:scope} use only
the following properties of the first-order interaction $f$: it is real,
phase invariant, superconformal and even under the reflection of
Section~\ref{sec:pairclosure}; its vector restriction is
$\res f=\res T_1$; and on the backgrounds of Section~\ref{sec:scope} its
independent-chiral scalar Hessian is $\Sigma\phi\bar\phi$, with all other
entries zero. Every interaction with these properties gives the same
zero-vev results, with $\Delta=f-T_1$ in \eqref{eq:deltastar}. The choice
\eqref{eq:additionalinteractions} uses only one additional primary density;
the zero-vev conditions do not select it uniquely. Agreement of the
quadratic actions on these backgrounds does not imply equivalence of
the full interacting theories. The higher-order phase-invariant
interactions are specified below.

\subsection{Completion with prescribed phase-invariant terms}
The construction of Section~\ref{sec:recursion} extends to any prescribed
sequence of real, local, phase-invariant functionals $C_n$, provided they
are defined on the same regular domain. In the induction of
Section~\ref{sec:induction}, replace
\eqref{ao:eq:induction} by
\begin{equation}
 H_n=C_n-\Pi_0K_n,\qquad
 \boxed{F_n=K_n-\Pi_0K_n+C_n.}
 \label{eq:prescribedinduction}
\end{equation}
The lower-order coefficients remain fixed, and the
phase-invariant auxiliary functional still generates a self-dual action.
At the first order where two solutions with the same prescribed
$C_n$ differ, their difference is phase invariant and has zero phase average, so uniqueness
holds with these data fixed. If every $C_n$ is superconformal, the same
covariance argument applies at each order. This establishes formal existence,
uniqueness and superconformality without a convergence assumption.

Let $F_n$ denote the canonical coefficients generated from $f_{\rm mod}$
and $T_n$ those generated from the minimal interaction. The higher-order
prescription uses the additional first-order term,
\begin{equation}
 \Delta=f_{\rm mod}-T_1=-\frac{152}{405}\,D_{33}.
 \label{eq:deltastar}
\end{equation}
It has zero vector restriction. At zero scalar expectation value its
independent-chiral scalar Hessian is $19\Sigma\phi\bar\phi$ in the
component normalization of Appendix~\ref{app:components}, and its other
entries vanish, including those involving the highest scalar components.
These properties are used in the recursive argument in
Appendix~\ref{app:components}.

The prescription in \eqref{eq:proposal} is
\begin{equation}
\begin{gathered}
 C_n=[t^n]d(t)\Delta,\qquad
 d(t)=-\frac{t^3}{76}+\frac{t^5}{152}-\frac{5t^7}{1216}+O(t^9),\\
 \widetilde F_1=f_{\rm mod},\qquad
 \widetilde F_2=F_2=\PP(f_{\rm mod},f_{\rm mod})+\cc,\qquad
 \widetilde F_3=F_3-\frac{\Delta}{76}.
\end{gathered}
\label{eq:neutralchoice}
\end{equation}
The prescribed terms are real, phase invariant, superconformal and
reflection even, with zero vector restriction. The closure argument in Section~\ref{sec:scalarpropagation} therefore
fixes the quadratic scalar series for the chosen $d(t)$. All charged
higher-order terms must be computed using
\eqref{eq:prescribedinduction}. Neither the zero-vev quadratic action
nor its restricted flow selects a unique full superspace interaction.

\section{Truncation to the \texorpdfstring{$\mathcal N=1$}{N=1} vector multiplet}
\label{sec:n1}
\subsection{Conventions and nilpotency identities}
We use $D^2=\tfrac12D^\alpha D_\alpha$ and normalize the Maxwell action as
\begin{equation}
 D^2=\tfrac12D^\alpha D_\alpha,\qquad
 W_\alpha=-\tfrac12\bar D^2D_\alpha V,\qquad
 \SM=\tfrac14\CI{W^2}+\cc,\qquad V=\bar V.
 \label{sr:eq:conventions}
\end{equation}
The relative full/chiral measure convention used for the $\mathcal N=1$
calculation is
\begin{equation}
 \FI{F}=-\tfrac12\CI{\bar D^2F},\qquad
 \delta\SM=\FI{E\,\delta V},\qquad E=D_\alpha W^\alpha.
 \label{sr:eq:projection}
\end{equation}
The conjugate chiral formula is understood. Define
\begin{equation}
 \begin{gathered}
 A=D^2W^2,\quad \bar A=\bar D^2\bar W^2,\quad N=W^2\bar W^2,\quad
 \Omega=\frac{4N}{A\bar A},\\
 u=D_\alpha W^\alpha,\quad v=\overline{D_\alpha W^\alpha},\qquad
 a=-\frac14(A+u^2/2),\quad b=\frac18u^2,
 \end{gathered}
 \label{sr:eq:composites}
\end{equation}
and use the conjugate definitions of $\bar a,\bar b$. The two chiral
arguments are independent when differentiating functionals. Only after
forming the recursive pairings do we impose the physical Bianchi
identity $u=v=E$. This distinction is relevant to divergence-dependent
actions, as discussed in~\cite{KS2023}.

In these conventions, the $\mathcal N=1$ ModMax action of~\cite[eq.~(4.19)]{BLST2021} reads
\begin{equation}
 S_{\rm B}(\gamma)=\cosh\gamma\,\SM+\sinh\gamma\,\RB,
 \qquad \RB=\FI{\frac{N}{\sqrt{A\bar A}}}.
 \label{sr:eq:target}
\end{equation}This is an off-shell superspace action. We shall show that the truncated
$\mathcal N=2$ action equals \eqref{sr:eq:target} after a local redefinition
of the real prepotential, without eliminating the auxiliary field.

Let $\mathscr I$ denote expressions containing at least one of the
undifferentiated spinors. Their nilpotency gives
\begin{equation}
 \mathscr I=(W_1,W_2,\bar W_{\dot1},\bar W_{\dot2}),\qquad
 N\mathscr I=0,\qquad W^2W_\alpha=0,\quad\bar W^2\bar W_{\dot\alpha}=0.
 \label{sr:eq:ideal}
\end{equation}Differentiation can remove the undifferentiated spinor factor. These
vanishing products therefore cannot be used before the derivatives act. Products with $N$ are therefore
always formed after the indicated differentiations, or after a chiral factor
$W^2$ has been taken through a barred derivative. In particular, $\partial_\mu
W_\alpha$, $D_\alpha W_\beta$ and $D^2W_\alpha$ are not set to zero.

\subsection{Reduction of the minimal interaction}
Choose $\theta=\theta_1$ and $\eta=\theta_2$. Switching off the $\mathcal N=1$ chiral multiplet sets $\Phi:=\W(x_{\rm ch},\theta,\eta=0)$ and all its components to zero. The remaining term in the $\eta$ expansion gives
\begin{equation}
 \W=2i\eta^\alpha W_\alpha,\qquad
 \res S_0=\SM.
 \label{sr:eq:embedding}
\end{equation}
The spacetime arguments in this formula are chiral coordinates.
The full extra chiral multiplet is removed; the retained vector
multiplet remains off shell. Let
$\rhoh$ denote evaluation at $\eta=\bar\eta=0$ after this embedding;
it does not set the surviving $\theta,\bar\theta$ to zero. 

The minimal interaction is
\begin{equation}
 T_1=2
 \int\dd^4x\,\dd^8\theta\,
 \frac{\W^2\bW^2}{Q},\qquad
 Q=\sqrt{\widehat K\cdot\widehat K},\quad
 \widehat K=-32X+128Y-32iZ-4T.
 \label{sr:eq:firstinteraction}
\end{equation}The coefficient 2 is the photon normalization for the full measure in
\eqref{eq:measures}. The bilinears are those in \eqref{eq:basis}, written
directly with $D^i_\alpha$ and $\bar D_{\dot\alpha i}$. The ordinary $SU(2)_R$
invariance makes different choices of the retained supersymmetry equivalent.

On the regular domain, with the branch fixed by the photon normalization,
the minimal interaction reduces to
\begin{equation}
 \boxed{\res T_1=\Rr=\FI{\frac{N}{R_1}},\qquad
 R_1=\sqrt{(A+u^2/2)(\bar A+v^2/2)},}
 \label{sr:eq:firstreduction}
\end{equation}
for arbitrary independent chiral spinors and before the Bianchi identity
is imposed. To see this, note that the numerator is of top degree in
$\eta$ and $\bar\eta$:
\begin{equation}
 \W^2\bW^2=16\eta^1\eta^2\bar\eta^{\dot1}\bar\eta^{\dot2}N.
 \label{sr:eq:hiddentop}
\end{equation}The $\eta$-dependent shifts of the chiral coordinates contain an additional
$\eta$ or $\bar\eta$ and therefore vanish in this product. Consequently the
integration over $\eta,\bar\eta$ involves only $\rhoh(Q^{-1})$ and no
$\eta$-derivatives of the denominator. With the Maxwell normalization of the
measures, the factor from the $\eta,\bar\eta$ integration in
\eqref{sr:eq:hiddentop} is 16. This is consistent with the chiral reduction
factor 2 in \eqref{sr:eq:embedding}, since our full $\mathcal N=2$ projector
is $(\bar D_1)^2(\bar D_2)^2$, whereas \eqref{sr:eq:projection} uses
$-\tfrac12\bar D^2$.

Write $f_{\alpha\beta}=D_{(\alpha}W_{\beta)}$. The structures $X,Y$
contain $\W$ or a spacetime derivative of its $\eta$-independent component and hence
vanish under $\rhoh$. The surviving terms of $Z$ contain
$W_\alpha\partial\bar W_{\dot\alpha}$ or
$(\partial W_\alpha)\bar W_{\dot\alpha}$, and are annihilated by $N$.
The mixed double-spinor derivatives in $T$ give the outer product
of $f$ and $\bar f$. Thus, with the index ordering of \eqref{eq:basis},
\begin{equation}
 N\rhoh(X)=N\rhoh(Y)=N\rhoh(Z)=0,\qquad
 N\rhoh(\widehat K_{\alpha\beta\dot\alpha\dot\beta})
 =32N f_{\alpha\beta}\bar f_{\dot\alpha\dot\beta}.
 \label{sr:eq:projectedtensor}
\end{equation}
An overall conjugation sign in the last relation would cancel in its
square. The spinor product rule, followed by multiplication by $N$, gives
\begin{equation}
 N f^{\alpha\beta}f_{\alpha\beta}=-N(A+u^2/2),\qquad
 N\rhoh(Q^2)=32^2 N(A+u^2/2)(\bar A+v^2/2).
 \label{sr:eq:rootnorm}
\end{equation}
For example, take $m_{\alpha\beta}=D_\alpha W_\beta$, with
$\varepsilon_{12}=1$. Then $u=m_{21}-m_{12}$ and
$NA=-2N\det m$; decomposing $m$ into its symmetric and antisymmetric
parts gives the first identity. The omitted terms in $A$ contain
$W^\alpha D^2W_\alpha$ and are annihilated by $N$.

All quantities in the last identity have already been differentiated.
Since their bodies are invertible, analytic functional calculus on the
chosen branch gives
\begin{equation}
 N\rhoh(Q^{-1})=\frac{N}{32R_1}.
 \label{sr:eq:rootinverse}
\end{equation}
Equations \eqref{sr:eq:firstinteraction}, \eqref{sr:eq:hiddentop} and
\eqref{sr:eq:rootinverse} give the overall factor $2\cdot16/32=1$, and
hence \eqref{sr:eq:firstreduction}.

\subsection{Truncation of the additional interaction}
\label{sec:modifiedrestriction}
For $\Phi=0$, one has $\W=2i\eta^\alpha W_\alpha$, so both
$\W^3$ and $\bW^3$, together with their derivatives, vanish identically.
The tensor $K_{33}$ in \eqref{eq:additionaldensities} is built from
$A=\W^3$ and $\bar B=\bW^3$ through \eqref{eq:mixedbasis}, so $D_{33}$
vanishes before integration. Therefore
\begin{equation}
 \res D_{33}=0,
 \qquad \res\Delta=0,\qquad\res f_{\rm mod}=\res T_1.
 \label{eq:modifiedrestriction}
\end{equation}
These are off-shell identities for independent chiral spinor arguments,
as required in the self-duality calculation. They allow arbitrary
spacetime dependence and do not use the vector Bianchi identity.

\subsection{Closure of the pairing of chiral functional derivatives}
\label{sec:pairclosure}
For a differentiable algebraic function $L(a,b,\bar a,\bar b)$ on the
regular domain define
\begin{equation}
 \FF[L]=\FI{\Omega L},\qquad f_L=\frac{4L}{A\bar A},\qquad
 r_{L\alpha}=\frac{\delta\FF[L]}{\delta W^\alpha}.
 \label{sr:eq:Fclass}
\end{equation}
The arbitrary superfields $W,\bar W$ in these expressions can carry
spacetime dependence and fermionic components. ``Algebraic'' only means
that $L$ is an ordinary function of the specified composites, not that
their component fields are constant.

For any two such functions $L,M$, the pairing of their chiral functional
derivatives closes on the same class:
\begin{equation}
 \boxed{\frac12\CI{r_L^\alpha r_{M\alpha}}
 =\FF\!\left[aL_aM_a+bL_bM_b\right].}
 \label{sr:eq:pairing}
\end{equation}
The two chiral arguments remain independent in this identity. To derive
\eqref{sr:eq:pairing}, we vary the full functional and integrate the
spinor derivatives by parts.
The exact chiral functional derivative of $\FI{Nf}$ is
\begin{equation}
 r_\alpha[f]=-\frac12\bar D^2
 \left\{2W_\alpha[\bar W^2 f+D^2(Nf_A)]+D_\alpha(Nf_u)\right\}.
 \label{sr:eq:exactgradient}
\end{equation}
Expanding the displayed $D$ derivatives first, and using
$D_\alpha\bar W_{\dot\alpha}=0$, put it in the form
\begin{equation}
 r_\alpha[f]=-\bar D^2\left\{\bar W^2
 \left[W_\alpha P_f+\tfrac12(D_\alpha W^2)Q_f\right]\right\}
 +W^2\mathcal X_\alpha[f],\qquad
 P_f=f+Af_A,\quad Q_f=f_u.
 \label{sr:eq:gradientfactorization}
\end{equation}Here $\mathcal X_\alpha$ collects the remaining local terms. They include
derivatives of $f_A,f_u$. This is an exact factorization: for example
$W_\alpha D_\beta W^2$ is proportional to $W^2D_\beta W_\alpha$, and $\bar D$
commutes through the chiral factor $W^2$. Thus all those terms retain an
undifferentiated factor $W^2$ even after the barred derivatives act. They do
not contribute to the product of two functional derivatives, because the other
derivative contains at least one undifferentiated factor $W$; no term is
discarded from an individual functional derivative.

The remaining expression carrying a spinor index is linear in $W$. In the component
index basis used for \eqref{sr:eq:rootnorm}, its coefficient matrix is
\begin{equation}
 C(P,Q)=\begin{pmatrix}
 P-m_{12}Q&m_{11}Q\\-m_{22}Q&P+m_{21}Q
 \end{pmatrix},\qquad
 W_\alpha P+\tfrac12(D_\alpha W^2)Q=C(P,Q)_\alpha{}^\beta W_\beta.
 \label{sr:eq:gradientmatrix}
\end{equation}After multiplying the two functional derivatives, convert one chiral projector
to a full integral using \eqref{sr:eq:projection}. The remaining projector is
multiplied by $\bar W^2$, so the identity\begin{equation}
 \bar W^2\bar D^2(\bar W^2 H)=\bar W^2\bar A H
 \label{sr:eq:barsaturation}
\end{equation}
applies for any even $H$. All terms where a barred derivative acts on $H$
still have an extra undifferentiated $\bar W$ and vanish. The polarized
determinant of \eqref{sr:eq:gradientmatrix}, together with
$N\det m=-NA/2$, therefore gives
\begin{equation}
 \frac12\CI{r[f]^\alpha r[g]_\alpha}
 =-\FI{N\bar A\left[
 P_fP_g+\frac u2(P_fQ_g+Q_fP_g)-\frac A2Q_fQ_g\right]}.
 \label{sr:eq:rawpair}
\end{equation}
This identity is applied only after all derivatives have acted, including
the mixed-derivative terms in
the barred projectors of \eqref{sr:eq:gradientfactorization}.

For $f=f_L$ the ordinary chain rule gives
\begin{equation}
 P_{f_L}=-\frac{L_a}{\bar A},\qquad
 Q_{f_L}=\frac{u(L_b-L_a)}{A\bar A}.
 \label{sr:eq:PQ}
\end{equation}
Substituting this and its $M$ counterpart into \eqref{sr:eq:rawpair} yields
\begin{equation}
 -\bar A\left[P_LP_M+\frac u2(P_LQ_M+Q_LP_M)-\frac A2Q_LQ_M\right]
 =\frac4{A\bar A}(aL_aM_a+bL_bM_b),
 \label{sr:eq:coefficientidentity}
\end{equation}
which gives \eqref{sr:eq:pairing}. The cancellation holds for arbitrary
$u$ and $v$.

The full $\mathcal N=2$ pairing also reduces correctly. The reflection
$\W(\eta)\mapsto-\W(-\eta)$ leaves the term linear in $\eta$ unchanged and
reverses the scalar superfields at orders $\eta^0$ and $\eta^2$. Both the
minimal interaction $T_1$ and $f_{\rm mod}$ are even under this reflection:
$K_{pq}$ acquires the factor $(-1)^{p+q}$ together with the reflection of
$\eta,\bar\eta$, so $\hK=-4K_{11}$ and $K_{33}$ are even. The
phase-invariant additions in \eqref{eq:neutralchoice} are even as well. The
contractions of functional derivatives and the phase averages defining every
coefficient preserve this reflection. At $\Phi=0$, which is invariant under this
reflection, the functional derivatives with respect to $\Phi$ and $\bar\Phi$
vanish. The chiral measure normalization then gives\begin{equation}
 \res\!\left(
 \int\dd^4x\,\dd^4\theta\,F_{\W}G_{\W}\right)
 =\frac12\CI{(\res F)_W^\alpha(\res G)_{W\alpha}}.
 \label{sr:eq:restrictionpairing}
\end{equation}Consequently \eqref{sr:eq:firstreduction} and \eqref{sr:eq:pairing}
apply inductively to the truncated canonical functionals.

\section{All-order equivalence with the off-shell \texorpdfstring{$\mathcal N=1$}{N=1} ModMax action}
\label{sec:matching}
Component field redefinitions have long been used to relate different
forms of supersymmetric nonlinear electrodynamics and Goldstino
actions~\cite{KM2005,KTy2011}. Here the change of variables is made at
the level of a real $\mathcal N=1$ prepotential, so that chirality and the
vector Bianchi identity follow from its definition. The result concerns
a formal off-shell equality on the regular domain and does not require
a Goldstino interpretation of the retained photino.

Our main result for the $\mathcal N=1$ truncation is the following. On the
regular domain specified below, the canonical completion of
\eqref{eq:minimalinteraction} admits a real, local, formal redefinition of the
$\mathcal N=1$ prepotential, $V_{\rm old}=V+O(t)$, such that
\begin{equation}
 \res S_{\sinh\gamma}[V_{\rm old}(V)]
 =\cosh\gamma\,\SM[V]+\sinh\gamma\,\RB[V]
 \label{eq:matchingidentity}
\end{equation}
holds to every order in $\gamma$, modulo superspace integration by parts
and with no equation of motion imposed on the retained $\mathcal N=1$
vector multiplet. The derivation has three steps. The exact truncation
identities close the canonical recursion on algebraic coefficient
functions. Their value at vanishing auxiliary invariant follows from uniqueness
of the canonical solution. The resulting identity ensures that the
prepotential redefinition remains regular at $E=0$.

\subsection{Truncated recursion at zero auxiliary invariant}
With the composites and functional $\FF$ of Section~\ref{sec:n1},
the truncated first-order interaction has density
\begin{equation}
 \res T_1=\FF[L_1],\qquad
 L_1=\frac{(a+b)(\bar a+\bar b)}{\sqrt{a\bar a}}.
 \label{eq:restrictedinteraction}
\end{equation}Let $\rho L=L(a,0,\bar a,0)$ denote evaluation of the coefficient functions at $b=\bar b=0$. This is distinct from the evaluation $\rhoh$ at
$\eta=\bar\eta=0$ used in the reduction of $T_1$.\begin{equation}
 \begin{aligned}
 \res\PP(F,G)&=\FF[p(L,M)],
 &p(L,M)&=aL_aM_a+bL_bM_b,\\
 \rho p(L,M)&=a(\rho L)_a(\rho M)_a,
 &\rho\Pi_0&=\Pi_0\rho,\qquad \rho L_1=\sqrt{a\bar a}.
 \end{aligned}
 \label{ao:eq:boundaryhomomorphism}
\end{equation}Here $\res F=\FF[L]$, $\res G=\FF[M]$ in the reflection-even class of
functionals generated by the first-order interaction. The reflection
$\W(\eta)\mapsto-\W(-\eta)$, which was used to derive the truncated pairing
\eqref{sr:eq:restrictionpairing}, is preserved by every pairing and every
phase average. Hence
the functional derivatives with respect to the omitted chiral multiplet vanish
at every order on $\Phi=0$. The same holds for the conjugate pairing.

The factors $b$ and $\bar b$ ensure that the pairing at $b=\bar b=0$
depends only on the coefficient functions at those values. The
functions and their derivatives are regular there, so no singular
terms survive multiplication by $b$ or $\bar b$. This evaluation is made on the
coefficient functions; no condition is imposed on the superfields.

Let $r_a=\sqrt{a\bar a}$, $C_t=\sqrt{1+t^2}$ and $k=C_t-1$.
At $b=\bar b=0$, consider the interaction
\begin{equation}
 \begin{gathered}
 J_\partial(t;a,\bar a)=\frac{k}{2}(a+\bar a)+t r_a,\\
 a(J_\partial)_a^2-\bar a(J_\partial)_{\bar a}^2
 =\frac{k^2-t^2}{4}(a-\bar a)
 =-\frac{k}{2}(a-\bar a).
 \end{gathered}
 \label{ao:eq:boundarysolution}
\end{equation}
Since $a$ and $b$ have $U(1)$ charge $+2$, the phase variation of
$J_\partial$ is $ik(a-\bar a)$. The last line of
\eqref{ao:eq:boundarysolution} therefore gives
$\delta_0J_\partial=4\im p(J_\partial,J_\partial)$ exactly.
Its order-one coefficient is $r_a$; all coefficients at higher orders
are proportional to $a+\bar a$ and have no phase-invariant part.
The same uniqueness argument applies to these coefficient functions:
at the first order where two solutions could differ, their difference
has zero phase variation and zero phase average, and hence vanishes. Together with
\eqref{ao:eq:boundaryhomomorphism}, this identifies $J_\partial$ with
the canonical solution evaluated at $b=\bar b=0$.

Once the Bianchi identity $u=v=E$ is imposed, write the truncated density
as $r\,\mathcal G(t;c,y)$, with $\mathcal G=c+y+\sum_{n\geq1}t^n\ell_n$.
The real variables $r,c,y$ are defined in \eqref{ao:eq:realvariables}.
The result is
\begin{equation}
 \begin{gathered}
 \boxed{\mathcal G(t;c,0)=c\sqrt{1+t^2}+t,}\\
 \ell_1(c,0)=1,\qquad
 \ell_{2j}(c,0)=\binom{1/2}{j}c,\qquad
 \ell_{2j+1}(c,0)=0\quad(j\geq1).
 \end{gathered}
 \label{ao:eq:allorderboundary}
\end{equation}
This determines the truncated coefficients at $y=0$ to every order.

\subsection{Construction of the prepotential redefinition}
Now impose $u=v=E$ and use a real prepotential. In our conventions
\begin{equation}
 \begin{gathered}
 W_\alpha=-\tfrac12\bar D^2D_\alpha V,\qquad
 E=D_\alpha W^\alpha,\quad A=D^2W^2,\quad
 N=W^2\bar W^2,\quad\Omega=\frac{4N}{A\bar A},\\
 x=-\frac{A+\bar A+E^2}{8},\qquad
 p=\frac{i(A-\bar A)}8,\qquad
 r=\sqrt{x^2+p^2},\qquad c=\frac xr,\qquad y=\frac{E^2}{8r}.
 \end{gathered}
 \label{ao:eq:realvariables}
\end{equation}These are composites, not new fields. On the domain considered, $r$, $A\bar A$
and the relevant square roots have invertible bodies. The branch near $y=0$
has $\varpi:=\sqrt{1+2cy+y^2}=1$ there and extends wherever these denominators
remain regular. The domain includes $E=0$; the construction uses no inverse
power of $E$. Statements near $y=0$ are local analytic statements in the
coefficient functions.

Homogeneity, reality, and the pairing closure imply that the truncated
canonical coefficients have densities $r\ell_n(c,y)$. More concretely,
the first-order interaction is polynomial in $a,b,\bar a,\bar b$ divided by a power of
$r_a$; the pairings and phase averages preserve this algebra.
Once $u=v=E$ is imposed, each coefficient is symmetric in $a,\bar a$, and the relations
$a\bar a=r^2$, $a+\bar a=2rc$ reduce it to a polynomial in $c,y$
at each order. In particular no coefficient develops an inverse power
of $y$.

At first order $\ell_1=1+2cy+y^2=\varpi^2$; the recursion based on the chiral pairing
gives $\ell_2=\tfrac12c(1+2cy+y^2)^2$. These two coefficients fix the
normalization of the matching calculation.

The Maxwell term and the square-root term of the target action can both be written as
\begin{equation}
 \SM=\FI{\Omega r(c+y)},\qquad
 \RB=\FI{\Omega r\varpi}.
 \label{eq:saturatedfree}
\end{equation}
Indeed $r(c+y)=-(A+\bar A)/8$. Since $\bar A$ is chiral,
$\bar D^2(\bar W^2/\bar A)=1$; the projection convention
\eqref{sr:eq:projection} and its conjugate then give the first identity.
The second follows from $A\bar A=16r^2\varpi^2$ on the chosen branch.

With $\SM=\tfrac14\int\dd^4x\,\dd^2\theta\,W^2+\cc$ and
$\RB=\int\dd^4x\,\dd^2\theta\,\dd^2\bar\theta\,
N/\sqrt{A\bar A}$, the two sides to be matched are
\begin{equation}
 \begin{gathered}
 \res S_t[V]=\FI{\Omega r\,\mathcal G(t;c,y)},\qquad
 \mathcal G(t;c,y)=c+y+\sum_{n\geq1}t^n\ell_n(c,y),\\
 S_{\rm B}(t)[V]=\FI{\Omega r\,\mathcal B(t;c,y)},\qquad
 \mathcal B(t;c,y)=\sqrt{1+t^2}(c+y)+t\varpi.
 \end{gathered}
 \label{ao:eq:densities}
\end{equation}The second line is the complete off-shell $\mathcal N=1$ ModMax action
of~\cite{BLST2021} in our conventions, including its fermionic and
auxiliary-field terms.

Seek a formal change of variable $Y=y+O(t)$ such that
\begin{equation}
 \mathcal G(t;c,Y(t;c,y))=\mathcal B(t;c,y),\qquad
 Y=y+\sum_{n\geq1}t^nq_n(c,y).
 \label{ao:eq:composition}
\end{equation}
The coefficient of $Y$ at $t=0$ is one, so each $q_n$ is determined
without dividing by a field-dependent Jacobian. To make the induction
explicit, let $Y_{<n}=y+\sum_{j=1}^{n-1}t^jq_j$, and define
$\beta_1=\varpi$, $\beta_{2j}=\binom{1/2}{j}(c+y)$ and
$\beta_{2j+1}=0$ for $j\geq1$. Coefficient comparison gives
\begin{equation}
 \boxed{q_n=\beta_n-
  \sum_{k=1}^{n}[t^{\,n-k}]\ell_k(c,Y_{<n}).}
 \label{ao:eq:qrecursion}
\end{equation}
This determines each order from the lower ones and constructs the formal
inverse around $Y=y$.

Suppose $q_j(c,0)=0$ below order $n$. At $y=0$ we have $Y_{<n}=0$;
only the $k=n$ term survives in the sum in \eqref{ao:eq:qrecursion}.
The identity \eqref{ao:eq:allorderboundary} gives
\begin{equation}
 \begin{gathered}
 q_n(c,0)=\beta_n(c,0)-\ell_n(c,0)=0,\qquad
 q_n=y\,d_n\quad\text{with }d_n\text{ regular},\\
 h=\sum_{n\geq1}t^nh_n,\qquad
 h_n=\frac12\left(d_n-\sum_{j=1}^{n-1}h_jh_{n-j}\right),\qquad
 Y=y(1+h)^2.
 \end{gathered}
 \label{ao:eq:regularmap}
\end{equation}The base case is the same identity with $n=1$. Since $q_n$ is analytic and vanishes at $y=0$, the ratio $d_n=q_n/y$
has a regular extension to $y=0$. The construction is therefore regular at a vanishing auxiliary
field. The square-root recursion for $h$ has leading coefficient $2$, hence is
uniquely solvable at every order.

Use the real, gauge-invariant prepotential shift
\begin{equation}
 \begin{gathered}
 \boxed{V_{\rm old}=V+\frac{NE}{A\bar A}h(t;c,y).}\\
 \Omega_{\rm old}=\Omega,\quad
 \Omega r_{\rm old}=\Omega r,\quad
 \Omega c_{\rm old}=\Omega c,\quad
 \Omega y_{\rm old}=\Omega Y.
 \end{gathered}
 \label{ao:eq:prepotential}
\end{equation}The last four identities follow from the finite superspace identities of
Appendix~\ref{app:finite}, valid for any local even $h$; their use does not
discard fermions or spacetime derivatives. Applied at each order, they remain
valid for the formal $h$ in \eqref{ao:eq:regularmap}. Since $h=O(t)$, the map
is locally invertible as a formal power series and preserves the prepotential
gauge transformation and the field strength Bianchi identity. At first order
the formula gives\begin{equation}
 q_1=\varpi-\varpi^2,\qquad
 d_1=-\frac{\varpi(2c+y)}{1+\varpi},\qquad
 h_1=-\frac{\varpi(2c+y)}{2(1+\varpi)},
 \label{ao:eq:mapfirst}
\end{equation}which is manifestly regular at $y=0$ on the branch $\varpi=1$. This supplies a
direct normalization check on the general induction.

Combining \eqref{ao:eq:composition} and \eqref{ao:eq:prepotential} gives
\begin{equation}
 \boxed{\begin{aligned}
 \res S_{\sinh\gamma}[V_{\rm old}(V)]
 ={}&\left(\frac14\cosh\gamma\int\dd^4x\,\dd^2\theta\,W^2+\cc\right)\\
 &+\sinh\gamma\int\dd^4x\,\dd^2\theta\,\dd^2\bar\theta\,
      \frac{W^2\bar W^2}{\sqrt{(D^2W^2)(\bar D^2\bar W^2)}}.
 \end{aligned}}\label{ao:eq:matching}
\end{equation}This equality holds to every order in $\gamma$, because $t=\sinh\gamma$ is an
invertible formal reparametrization at zero. The photon, photino and real auxiliary field
of the vector multiplet remain off shell. The additional $\mathcal N=1$ chiral
multiplet has been removed by $\res$, as specified in Section~\ref{sec:n1}.
This establishes \eqref{eq:matchingidentity} for the canonical completion
of $T_1$.

\subsection{Equivalence for the modified completion}
The equivalence \eqref{eq:matchingidentity} was established for the
canonical completion of $T_1$, namely $S_t=S_0+\sum t^nT_n$. It extends
to the action \eqref{eq:proposal} by the following induction.
Equation~\eqref{eq:modifiedrestriction} gives the first-order equality.
Suppose $\res\widetilde F_k=\res T_k$ for every $k<n$.
Reflection invariance and \eqref{sr:eq:restrictionpairing} identify the
truncated source of every chiral pairing in the two order-$n$
self-duality equations. Hence
\begin{equation}
 \delta_0\res(\widetilde F_n-T_n)=0,\qquad
 \Pi_0\res(\widetilde F_n-T_n)=\res C_n=0.
 \label{eq:modifiedmatchinginduction}
\end{equation}
The first equality says the difference is phase-invariant; the second then makes
it zero. The same argument with $C_n=0$ applies to the canonical modified
family $F_n$. Therefore
\begin{equation}
 \boxed{\res\widetilde F_n=\res F_n=\res T_n\quad(n\geq1),\qquad
 \res\widetilde S_{\sinh\gamma}[V]
       =S_{\rm B}(\gamma;V_{\rm B}[V]).}
 \label{eq:modifiedmatching}
\end{equation}The same $\mathcal N=1$ prepotential redefinition therefore establishes the
equivalence for the modified completion.

\section{Physical interpretation and propagation}
\label{sec:scope}
\subsection{Scope of the propagation calculation}
The following results concern quadratic fluctuations of
\eqref{eq:proposal} on constant non-null electromagnetic backgrounds
with vanishing scalar expectation value. The quadratic action is determined to all orders in the coupling, and
its coefficient functions are analytically continued. The additional chiral
multiplet is retained: zero scalar vev is not the restriction
$\Phi=0$ used for the $\mathcal N=1$ equivalence. No propagation
claim on other backgrounds is made here. In particular, propagation at
nonzero scalar expectation value lies outside the scope of this paper.

\subsection{Scalar propagation at vanishing scalar expectation value}
\label{sec:scalarpropagation}
Take a constant non-null electromagnetic field and vanishing scalar,
fermion and auxiliary backgrounds. A frame with parallel electric and
magnetic fields covers the whole class:
\begin{equation}
 \mathbf E=E_0\mathbf e_2,\qquad \mathbf B=B_0\mathbf e_2,\qquad
 r=\frac{E_0^2+B_0^2}{2}>0,\qquad z_0=d_0^a=\psi_0^i=0.
 \label{eq:constantbackground}
\end{equation}
These configurations solve the equations at every formal order.
The gauge equation is the divergence of a constant constitutive tensor.
The background is fixed by the reflection of
Section~\ref{sec:pairclosure}, so the action contains no terms linear in the scalar fluctuations.
It is also an $SU(2)_R$ singlet, excluding a term linear in the
auxiliary triplet, while fermion parity excludes terms linear in the fermion
fluctuations. These statements are preserved by the recursion and the
prescribed phase-invariant interactions.

Scalar--photon mixing is absent in the quadratic action on this class. The
discrete reflection of Section~\ref{sec:pairclosure} reverses the scalar components
and leaves the vector components unchanged; the background is a fixed point.
Equivalently, unbroken $U(1)_R$ forbids a scalar--photon bilinear. The scalar
is an $SU(2)_R$ singlet whereas the auxiliary fields form a triplet, excluding
their mixing. Fermion parity separates bosons and fermions. These selection
rules are preserved by the full recursion and the prescribed phase-invariant
coefficients. The scalar Hessian computed below therefore describes a
decoupled sector.

Appendix~\ref{app:components} computes the independent-chiral scalar
Hessian at arbitrary momentum and establishes closure under the recursion.
For the canonical modified completion, all higher scalar Hessians vanish:
$(F_n)^{(2)}_z=0$ for $n\geq2$. Its normalized quadratic density is
\begin{equation}
 L_{z,\mathrm{can}}^{(2)}
 =-\left(\eta^{\mu\nu}-\frac{t}{2r}T_F^{\mu\nu}\right)
             \partial_\mu z\partial_\nu\bar z,
 \qquad
 v_{\perp,\mathrm{can}}^2=\frac{1-t/2}{1+t/2}.
 \label{eq:canonicalscalar}
\end{equation}
Here $T_F$ is the Maxwell stress tensor of the background. After analytic continuation in $t$, the coefficient of the transverse
gradient term vanishes at $t=2$ and becomes negative for $t>2$, giving a
gradient instability. This is the reason for specifying
higher phase-invariant terms rather than adopting the canonical prescription.

The scalar Hessian of $\Delta$ is nineteen times that of $f_{\rm mod}$.
For the action \eqref{eq:proposal}, the same closure argument gives\begin{equation}
 \begin{gathered}
 (\widetilde F_n)^{(2)}_z=[t^n]d(t)\,(\Delta)^{(2)}_z\quad(n\geq2),\\
 \beta(t)=\frac{t+19d(t)}2
          =\frac{t}{\sqrt{1+t^2}+1}=\tanh\frac\gamma2,\\
 \boxed{L_z^{(2)}=-\left(\eta^{\mu\nu}
                -\frac{\beta(t)}rT_F^{\mu\nu}\right)
                \partial_\mu z\partial_\nu\bar z.}
 \end{gathered}
 \label{eq:scalarquadratic}
\end{equation}Within this quadratic form, positive kinetic and spatial coefficients
and subluminality hold if and only if $0\leq\beta<1$; these inequalities alone do
not determine $\beta(t)$. The choice above is selected by the restricted
root-flow condition of Section~\ref{sec:rootflow}. It fixes every
coefficient of the scalar quadratic series. The closed form
of $\beta(t)$ defines its analytic continuation along the positive real $t$
axis.

In the frame \eqref{eq:constantbackground}, with contravariant indices,
\begin{equation}
 \frac{T_F^{\mu\nu}}r=\operatorname{diag}(1,1,-1,1),\qquad
 K_z=(1+\beta)(\omega^2-q_\parallel^2)-(1-\beta)q_\perp^2.
 \label{eq:scalaroperator}
\end{equation}The negative spatial entry reflects the Maxwell tension along the field
direction $x^2$; the time component is positive. Writing $q_\parallel=q_2$ and
$q_\perp^2=q_1^2+q_3^2$, one obtains\begin{equation}
 \boxed{\omega^2=q_\parallel^2+e^{-\gamma}q_\perp^2,\qquad
 v_\parallel^2=1,\qquad v_\perp^2=e^{-\gamma},\qquad
 1\pm\beta>0\quad(0\leq\gamma<\infty).}
 \label{eq:scalardispersion}
\end{equation}
Both real scalar components obey this dispersion relation. For a general function
$d(t)$, \eqref{eq:scalaroperator} gives $v_\perp^2=(1-\beta)/(1+\beta)$
with $\beta=(t+19d(t))/2$. The positivity, hyperbolicity and causality
statements below therefore hold for every $d(t)$ with $0\leq\beta(t)<1$;
the root-flow choice of Section~\ref{sec:rootflow} selects the value
$e^{-\gamma}$. The photon, gaugino and auxiliary operators of
Section~\ref{sec:zeromultiplet} follow from the $\mathcal N=1$
equivalence and $SU(2)_R$ and do not depend on $d(t)$. The positive time
and spatial coefficients give a positive quadratic Hamiltonian and a
hyperbolic scalar equation. In the same frame the ray speed satisfies
\begin{equation}
 |\mathbf v_g|^2
 =\frac{q_\parallel^2+e^{-2\gamma}q_\perp^2}
        {q_\parallel^2+e^{-\gamma}q_\perp^2}\leq1.
 \label{eq:scalarrays}
\end{equation}Hence the scalar characteristic cone lies inside or on the Minkowski light
cone. Because the quadratic operator is exactly second order in derivatives,
its characteristic cone is obtained directly, rather than from the
low-frequency limit of a truncated dispersive operator. The interpretation in
terms of an effective scalar metric is the usual one for quadratic
fluctuations~\cite{BMV2008}, while the distinction between characteristic
cones and well-posed evolution is discussed in~\cite{ACGR2015,Reall2021}. The
limit $\gamma\to\infty$ is degenerate because the transverse coefficient tends
to zero.

In the same parallel-field frame, the bosonic ModMax photon cones are~\cite{BLST2020}
\begin{equation}
 \omega_+^2=q_\parallel^2+q_\perp^2,\qquad
 \omega_-^2=q_\parallel^2+e^{-2\gamma}q_\perp^2.
 \label{eq:photoncones}
\end{equation}The scalar cone lies between these two cones for positive coupling. The
background need not preserve supersymmetry, so the scalar and photon cones
need not coincide. The remaining quadratic sectors can be determined
using the linearization of the vector-prepotential redefinition, as
we now explain.

\subsection{The remaining fields at vanishing scalar expectation value}
\label{sec:zeromultiplet}
The $\mathcal N=1$ equivalence also determines the photon and photino
quadratic operators on \eqref{eq:constantbackground}. The reflection
used in Section~\ref{sec:pairclosure} excludes mixing with $\Phi$.
Moreover, the linearization of the prepotential change
\eqref{sr:eq:Vshift} is particularly simple here. The background has
$E_0=0$, while $N_0=W_0^2\bar W_0^2$ already contains
$\theta^2\bar\theta^2$. Thus the linearized change is proportional
to $N_0\delta E$: it changes only the auxiliary component of the
real prepotential. It is the identity on the photon and photino.
Their quadratic operators in the original variables therefore agree
with those of the Bandos--Lechner--Sorokin--Townsend action.

On a constant electromagnetic background the term proportional to the
derivative of the electromagnetic phase in Eq.~(4.28) of~\cite{BLST2021}
vanishes. Relative to the free Weyl operator, the photino coefficient is
\begin{equation}
 Z_\lambda=\cosh\gamma+\frac{x}{r}\sinh\gamma
 \geq e^{-\gamma}>0\qquad(0\leq\gamma<\infty).
 \label{eq:zerogaugino}
\end{equation}
The background preserves $SU(2)_R$, so both gauginos have this operator.
There is no second independent mixed-chirality coefficient, and
$U(1)_R$ excludes same-chirality bilinears. Both gauginos consequently
have the Minkowski light cone. This conclusion uses the quadratic
linearization of the equivalence, as well as the background symmetries;
equivalence of the full formal actions alone would not establish it.

The auxiliary coefficient can be obtained directly from the restricted
recursion. In the notation of Section~\ref{sec:matching}, expand its
interaction density as
\begin{equation}
 J=J_\partial+b\,u(\vartheta,t)+\bar b\,\overline{u(\vartheta,t)}
       +O(b^2,b\bar b,\bar b^2),\qquad
 \upsilon=\frac{a}{\sqrt{a\bar a}}=e^{i\vartheta},\quad
 \cos\vartheta=\frac xr.
 \label{eq:auxexpansion}
\end{equation}
Here $u$ is a coefficient function, distinct from the superspace
composite $u$ of \eqref{sr:eq:composites}. Equating the terms linear
in $b$ in $\delta_0J=4\im p(J,J)$ gives
\begin{equation}
 -i\bigl(\sqrt{1+t^2}+t\cos\vartheta\bigr)\partial_\vartheta u
       +u+u^2=0,\qquad [\upsilon^{-1}]u=t.
 \label{eq:auxode}
\end{equation}
The last condition is the prescribed phase average: $b$ has charge
$+2$, so the $\upsilon^{-1}$ coefficient is fixed by the first-order
interaction. Put $\tau=\tanh(\gamma/2)$. The solution connected to
$u=0$ at $t=0$ is
\begin{equation}
 u=\frac{w}{1-w},\qquad
 w=\rho\frac{1+\tau\upsilon}{\upsilon+\tau},\qquad
 \frac{\rho(1-\tau^2)}{(1-\rho\tau)^2}=t.
 \label{eq:auxsolution}
\end{equation}
Substitution verifies the differential equation. The Laurent expansion
of $u$ about $|\upsilon|=1$ gives the last relation from
$[\upsilon^{-1}]u=t$. For $0<\tau<1$, the function
$\rho(1-\tau^2)^2-2\tau(1-\rho\tau)^2$ is strictly increasing on
$[0,1]$, negative at zero and positive at one. Hence there is one
solution with $0<\rho<1$. Since $|w|=\rho$ on the unit circle, the
auxiliary coefficient, relative to its free value, is
\begin{equation}
 Z_d=1+u+\bar u=\frac{1-\rho^2}{|1-w|^2}>0,
 \qquad
 Z_d=1+2\frac xr t+2\frac{x^2}{r^2}t^2+O(t^3).
 \label{eq:zeroaux}
\end{equation}
No spacetime derivative acts on the auxiliary fluctuation in this
quadratic action. $SU(2)_R$ gives the same coefficient for all three
real auxiliary fields.

Together with \eqref{eq:scalardispersion} and \eqref{eq:photoncones},
these results determine the quadratic series of the whole multiplet
on \eqref{eq:constantbackground}. Its analytic continuation has
causal scalar, photon and gaugino propagation and algebraic auxiliary
fields at every finite $\gamma\geq0$.

\subsection{Comparison with compensator constructions}
\label{sec:compensators}
Kuzenko and Raptakis, Section~4.3 of~\cite{KR2021}, note that no $\mathcal N=2$ superconformal ModMax theory had been constructed
and identify the dimensionless full superspace measure as the obstacle. Equation numbers in this comparison refer to arXiv version~5. Their
Eq.~(4.28) supplies a compensator-dependent $\mathcal N=2$ model using
an auxiliary chiral scalar and a compensating vector field strength.
After switching off the additional physical $\mathcal N=1$ chiral multiplet and eliminating the auxiliary chiral components, their model reduces to $\mathcal N=1$ ModMax; see their
Eqs.~(4.29)--(4.32). An $\mathcal N=2$ model whose truncation is
$\mathcal N=1$ ModMax therefore already exists in the literature. The present construction uses no additional compensator in the
$\mathcal N=2$ action.

For the minimal interaction $T_1$ the relevant dimensions are
\begin{equation}
 [\dd^4x\,\dd^8\theta]=0,\qquad
 [\W^2\bW^2]=[Q]=4,\qquad
 \left[\frac{\W^2\bW^2}{Q}\right]=0.
 \label{eq:dimensionbalance}
\end{equation}
The derivative composite $Q$ supplies the required weight, and the
additional density $D_{33}$ in \eqref{eq:additionalinteractions} also has a dimensionless
superspace integrand. Both are constructed using only
the physical field strength. Dimension counting alone is insufficient:
Section~\ref{sec:primary} establishes the cancellation of its inhomogeneous
superconformal variation. This choice admits $z_0=0$ at nonzero
electromagnetic invariants, at the cost of a nonanalytic denominator
and a derivative-dependent interaction on the regular domain specified in subsection ~3.4.

Section~4 of ref.~\cite{K2021} noted that the $\mathcal N=1$ superconformal
ModMax model does not appear to have an $\mathcal N=2$ counterpart.
Superconformal duality-invariant $\mathcal N=2$ models are constructed
in~\cite{K2021} itself, and general methods are reviewed in~\cite{K2026}. Those models involve logarithms and inverse powers of $\W$ and are
defined on a domain where its lowest scalar component is nonzero.
The present interactions instead admit $z_0=0$ on constant non-null
electromagnetic backgrounds, subject to the regularity conditions
specified in Section~3.4.

Higher-derivative ModMax deformations preserving duality and Weyl invariance
are available in the bosonic and $\mathcal N=1$ supersymmetric
settings~\cite{KR2024}. They demonstrate that the symmetry requirements allow
interactions beyond the original derivative-free electromagnetic Lagrangian.
They do not determine the particular first-order interaction or choice of the
phase-invariant terms used here. The additional primary density in
\eqref{eq:additionalinteractions} alters the first-order scalar Hessian while preserving
$\res f_{\rm mod}=\res T_1$; the higher phase-invariant coefficients then
determine the scalar quadratic action at higher orders.

\subsection{Comparison with higher-spin currents}
\label{sec:currentscomparison}
Off-shell cubic couplings of the $\mathcal N=2$ hypermultiplet
to higher-spin gauge superfields and the associated conserved
supercurrents were constructed in harmonic superspace by
Buchbinder, Ivanov and Zaigraev~\cite{BIZ2022,BIZ2023}.
Kuzenko and Raptakis studied conformal and non-conformal
$\mathcal N=2$ higher-spin supercurrent multiplets, including
conformal currents built from the vector multiplet~\cite{KR2023}.
Higher-spin currents and symmetries of the vector multiplet
were studied further in~\cite{Z2026}.

The vector-multiplet currents $J_{\alpha(s)\dot\alpha(s)}$
of~\cite{KR2023} are conserved when the field strengths obey
the free equations of motion. Their Eqs.~(2.10)--(2.12)
contain the relevant $s=2$ bilinear sector. Its first three
structures have the ratios $-X+4Y-iZ$, also present in
$\hK/32$. The remaining structure uses contracted two-spinor
derivatives. A complete identification with that current,
including conjugate derivative ordering and possible
equation-of-motion or improvement terms, has not been
established. We do not claim novelty for the bilinear
current itself.

Our construction uses its off-shell transformation law rather than a
conservation equation valid on the free equations of motion. A primary superfield with a spin-two
lowest component need not be the stress tensor defined by metric variation,
and we do not identify $\hK$ with that tensor. The
$\mathcal N=2$ supercurrent multiplet corresponds to a scalar superfield
in~\cite{KR2023} and couples to supergravity~\cite{BK2010}. Primarity
alone therefore implies neither causal propagation nor a stress-tensor
flow equation.

\subsection{Comparison with stress-tensor flows}
\label{sec:rootflow}
Stress-tensor flows provide a separate way of constructing
duality-preserving deformations~\cite{BBYM2022,FST2022,FST2023,FKST2023,BHO2024}.
We compare the component action with
\begin{equation}
 \partial_\gamma L=R[T],\qquad
 R[T]=\frac12\sqrt{T_{\mu\nu}T^{\mu\nu}
                    -\frac14(T_\mu{}^\mu)^2}.
 \label{eq:stressrootdefinition}
\end{equation}
The four-dimensional square-root operator was introduced by
Babaei-Aghbolagh et al. in Eq.~(3.6) of~\cite{BBYM2022}; their
Eqs.~(3.5) and (4.1) give the normalization and ModMax flow used here.
The choice of stress tensor matters. At zero scalar expectation value,
we vary the photon and scalar quadratic densities with the metric,
holding $F_{\mu\nu}$ and the scalar fixed. A conformal scalar
improvement changes the quadratic comparison only by a total
derivative on a constant electromagnetic background. All scalar
comparisons in this subsection are restricted to quadratic fluctuations
at vanishing scalar expectation value.

For ModMax, $T_{0\mu\nu}=\kappa_\gamma T_{F\mu\nu}$, with
$\kappa_\gamma=\cosh\gamma+(x/r)\sinh\gamma>0$. Since
$T_{F\mu\nu}T_F^{\mu\nu}=4r^2$, the positive square root gives
$R_0=\kappa_\gamma r=\partial_\gamma L_{\rm MM}$.
For the scalar quadratic density at $z_0=0$, put
\begin{equation}
 M^{\mu\nu}=T_F^{\mu\nu}/r,\qquad
 s_z=\partial_\mu z\partial^\mu\bar z,\qquad
 u_z=M^{\mu\nu}\partial_\mu z\partial_\nu\bar z,\qquad
 L_z^{(2)}=-s_z+\beta u_z.
 \label{eq:scalarflowvariables}
\end{equation}
The tensor $M$ must be varied along with the metric. The directional
variation $\delta_M g^{\mu\nu}=M^{\mu\nu}$ gives
$\delta_M s_z=u_z$ and $\delta_M u_z=s_z$. Expanding the root yields
\begin{equation}
 R^{(2)}=\frac12(u_z-\beta s_z),\qquad
 R^{(2)}-\partial_\gamma L_z^{(2)}-\frac\beta2L_z^{(2)}
 =\left[\frac12(1-\beta^2)-\partial_\gamma\beta\right]u_z.
 \label{eq:scalarflowresidual}
\end{equation}
With the Maxwell initial condition $\beta(0)=0$, the vanishing of the
bracket gives $\partial_\gamma\beta=(1-\beta^2)/2$ and hence
$\beta=\tanh(\gamma/2)$. Through $\beta=(t+19d(t))/2$, this fixes the
function $d(t)$ used in \eqref{eq:proposal}. The remaining term
is proportional to the quadratic scalar action and hence to its
equations of motion modulo a divergence. Equivalently, the change of
normalization
\begin{equation}
 z=\sqrt{\cosh(\gamma/2)}\,\widehat z,\qquad
 \widehat L_z^{(2)}=-\cosh(\gamma/2)\widehat s_z
                     +\sinh(\gamma/2)\widehat u_z
 \label{eq:scalarflownormalization}
\end{equation}
gives $\partial_\gamma\widehat L_z^{(2)}=\widehat R^{(2)}$ directly.
The scalar characteristic cone is unchanged.

The photon and quadratic scalar sectors therefore satisfy the stated
component root flow on the backgrounds considered here. This agreement
fixes the chosen scalar coefficient function but does not establish a
supersymmetric flow for the full action. The construction of an
interacting conserved supercurrent and the associated flow lies outside
the scope of this paper. The action \eqref{eq:proposal} is defined by
self-duality and the prescribed phase-invariant interactions.

\section{Conclusions}
The prescribed action \eqref{eq:proposal} is a compensator-free,
superconformal and self-dual formal series for an Abelian $\mathcal N=2$
vector multiplet. Its $\mathcal N=1$ vector restriction is equivalent
off shell to the Bandos--Lechner--Sorokin--Townsend action through a
local prepotential change. The construction is defined on a regular
domain where the composite field-dependent denominator is nonzero and does not include constant zero-field vacua.

The scalar dynamics requires additional choices beyond symmetry and
the vector restriction. On constant non-null electromagnetic backgrounds
with vanishing scalar expectation value, the prescribed phase-invariant
terms determine the quadratic action to all orders. Its analytic
continuation has scalar transverse speed squared $e^{-\gamma}$,
the ModMax photon cones, luminal gauginos and algebraic auxiliaries,
with positive kinetic coefficients for finite $\gamma\geq0$. Supersymmetry of the action alone does not imply that all component
fields have the same characteristic cone on a given electromagnetic
background.

Within the quadratic scalar form studied here, the stress-tensor
square-root flow modulo equations of motion and a total derivative
fixes the function $d(t)$. This is a condition on that sector, not
a derivation of the full action from a supersymmetric flow. Causal
scalar propagation holds for every $d(t)$ with $0\leq\beta<1$; the
flow selects $v_\perp^2=e^{-\gamma}$. An interacting conserved
supercurrent and a supersymmetric flow generating the full action have
not been established.

The superconformal and self-duality requirements, the vector restriction,
and the zero-vev propagation conditions leave freedom in the full
superspace interactions. The zero-vev results hold for a class of
first-order interactions, of which \eqref{eq:additionalinteractions} is
a simple representative. Propagation beyond the backgrounds considered here remain open. The analysis at nonzero scalar expectation value and the construction
of a supersymmetric flow in terms of a conserved supercurrent will
be addressed in a separate paper.

\appendix
\section{Finite identities for the prepotential redefinition}
\label{app:finite}
We now impose the Bianchi identity $u=v=E$ and work with a real
prepotential. For any real local even $h$, consider the finite shift
\begin{equation}
 V'=V+NH,\qquad H=\frac{E}{A\bar A}h,\qquad
 W'_\alpha=W_\alpha-\tfrac12\bar D^2D_\alpha(NH).
 \label{sr:eq:Vshift}
\end{equation}
In the application $h=O(t)$, so the transformation is formally
invertible near the identity and preserves the gauge transformation of
$V$. The corresponding $W'$ is a chiral field strength satisfying its
off-shell Bianchi identity.

Define
\begin{equation}
 \delta=\frac{E^2}{8},\quad x=-\frac{A+\bar A+E^2}{8},\quad
 p=\frac{i(A-\bar A)}8,\quad r=\sqrt{x^2+p^2}.
 \label{sr:eq:realcoordinates}
\end{equation}
On the regular domain, the finite shift \eqref{sr:eq:Vshift} satisfies the
superfield identities
\begin{equation}
 \boxed{\Omega'=\Omega,\quad \Omega x'=\Omega x,\quad
 \Omega p'=\Omega p,\quad\Omega E'=\Omega E(1+h),\quad
 \Omega\delta'=\Omega\delta(1+h)^2.}
 \label{sr:eq:finiteshift}
\end{equation}
The factors of $\Omega$ are part of these identities and must be retained.
To derive them, we expand the $D$ derivative in \eqref{sr:eq:Vshift}:
\begin{equation}
 W'_\alpha=W_\alpha-\tfrac12\bar D^2
 \{\bar W^2(D_\alpha W^2)H\}+W^2\mathcal Y_\alpha
 =C_\alpha{}^\beta W_\beta+W^2\mathcal Y_\alpha.
 \label{sr:eq:Wprimefactor}
\end{equation}
All derivatives of $H$ in the omitted expression remain in the
explicit $W^2\mathcal Y_\alpha$ remainder or in the chiral matrix $C$.
The factorization is exact. The nilpotency of the Grassmann-odd spinor superfields implies
\begin{equation}
 W'^2=(\det C)W^2,\qquad
 N'=N\det C\det\bar C.
 \label{sr:eq:numeratordeterminant}
\end{equation}
Differentiate the exact expression for $W'^2$ before multiplying by $N$. The product rule gives
\begin{equation}
 NA'=NA\det C,\qquad N\bar A'=N\bar A\det\bar C.
 \label{sr:eq:denominatordeterminant}
\end{equation}
Every other term has an undifferentiated factor $W$ or $\bar W$ and vanishes when multiplied by $N$.
Invertibility of the bodies then gives the \emph{finite} identity
\begin{equation}
 \frac{4N'}{A'\bar A'}
 =\frac{4N\det C\det\bar C}{A\bar A\det C\det\bar C}
 =\Omega.
 \label{sr:eq:Omegaderivation}
\end{equation}

For the remaining identities compute $D_\alpha W'_\beta$ from the exact
factorization \eqref{sr:eq:Wprimefactor}. Terms in which $D$ acts on $C$,
or the $W^2\mathcal Y$ remainder, retain an undifferentiated factor $W$ and vanish
after multiplication by $N$. Evaluating the surviving matrix with
\eqref{sr:eq:barsaturation} gives
\begin{equation}
 N\delta m_{\alpha\beta}
 =-\frac12 N A\bar A H\,\varepsilon_{\alpha\beta}
 =-\frac12 N E h\,\varepsilon_{\alpha\beta}.
 \label{sr:eq:traceshift}
\end{equation}
This can be seen explicitly from
\begin{equation}
 N(C-\mathbf1)=N\bar A H
 \begin{pmatrix}m_{12}&-m_{11}\\m_{22}&-m_{21}\end{pmatrix},\qquad
 N m'=N m C^{\mathsf T}.
 \label{sr:eq:shiftmatrix}
\end{equation}
The symmetric part of $m$ is unchanged, whereas
$N E'=N E(1+h)$. Applying the exact product rule for $A'$ and the
spinor determinant identity already used in \eqref{sr:eq:rootnorm}, one finds
\begin{equation}
 N\left(A'+\tfrac12E'^2\right)=N\left(A+\tfrac12E^2\right),\qquad
 N\left(\bar A'+\tfrac12E'^2\right)
 =N\left(\bar A+\tfrac12E^2\right).
 \label{sr:eq:tracelessinvariants}
\end{equation}Together with \eqref{sr:eq:Omegaderivation}, these equations give all the
identities in \eqref{sr:eq:finiteshift}. The same identities hold for analytic
functions of these invariants after multiplication by $\Omega$, on the chosen
branch.

\section{Scalar quadratic action and recursive closure}
\label{app:components}
We calculate scalar fluctuations about \eqref{eq:constantbackground},
retaining the additional chiral multiplet and arbitrary momentum.
Independent chiral arguments must be varied before imposing reduced
chirality, as required by the self-duality equation.

\subsection{Conventions and background normalization}
In chiral coordinates use
\begin{equation}
 D^i_\alpha=\frac{\partial}{\partial\theta_i^\alpha}
       +2i\bar\theta^{\dot\beta i}\partial_{\alpha\dot\beta},\qquad
 \bar D_{\dot\alpha i}=-\frac{\partial}{\partial\bar\theta^{\dot\alpha i}}.
 \label{eq:componentoperators}
\end{equation}
The momentum matrix and ordered monomials are
\begin{equation}
 \begin{gathered}
 p_{\alpha\dot\alpha}=
  \begin{pmatrix}\omega+q_3&q_1-iq_2\\q_1+iq_2&\omega-q_3\end{pmatrix},
 \quad P=\det p=\omega^2-q_1^2-q_2^2-q_3^2,\\
 \Theta=\theta_1^1\theta_2^1+\theta_1^2\theta_2^2,\quad
 \bar\Theta=\bar\theta^{\dot1 1}\bar\theta^{\dot1 2}
            +\bar\theta^{\dot2 1}\bar\theta^{\dot2 2},\\
 \theta^4=\theta_1^1\theta_1^2\theta_2^1\theta_2^2,\quad
 \bar\theta^4=\bar\theta^{\dot1 1}\bar\theta^{\dot2 1}
                         \bar\theta^{\dot1 2}\bar\theta^{\dot2 2}.
 \end{gathered}
 \label{eq:offshellmomentum}
\end{equation}
For parallel fields along $x^2$ choose
\begin{equation}
 \W_F=-f\Theta,\qquad \bW_F=g\bar\Theta,\qquad
 f=b_F+ie_F,\quad g=b_F-ie_F,\quad b_F,e_F\in\mathbb R.
 \label{eq:componentbackground}
\end{equation}We may keep $f,g$ algebraically independent during extraction and impose
reality afterwards. We write the chiral Grassmann integral as $C_C>0$ times the
coefficient of the ordered monomial $\theta^4$. Component coefficients are
converted to the normalization \eqref{eq:measures} before constants are
compared. The relation to the physical fields is\begin{equation}
 \begin{gathered}
 B_0=\sqrt{C_C}\,b_F,\qquad E_0=\sqrt{C_C}\,e_F,\\
 x=-\frac{C_C}{4}(f^2+g^2),\quad p=C_Cb_Fe_F,\quad
 r=\frac{C_C}{2}fg>0,\quad
 \nu:=\frac{f^2+g^2}{2fg}=-\frac{x}{r}.
 \end{gathered}
 \label{eq:componentdictionary}
\end{equation}
The orientation is chosen so that $p=\mathbf E\cdot\mathbf B$.
Reversing that convention leaves the quadratic results unchanged.
The normalization of \eqref{eq:calibration} gives
\begin{equation}
 \begin{gathered}
 L_1=2C_C\left[\frac{N}{Q}\right]_{\theta^4\bar\theta^4},\qquad
 N=\W^2\bW^2,\quad Q_0=16fg,\quad H=\hK^2-Q_0^2,\\
 L_{0,F}=-\frac{C_C}{4}(f^2+g^2)=x,\qquad
 L_{1,F}=\frac{C_C}{2}fg=r,\\
 L_{0,\phi}^{(2)}=-2C_C\partial\phi\cdot\partial\bar\phi,
 \qquad z=\sqrt{2C_C}\,\phi.
 \end{gathered}
 \label{eq:componentmeasure}
\end{equation}
The positive branch of the square root in the photon sector and the standard sign of the free scalar
kinetic term therefore follow from the same normalization. The purely magnetic specialization
is $f=g\ne0$, with $B_0^2=C_Cf^2$.

\subsection{Scalar Hessian before imposing reduced chirality}
Let $\mathcal T$ denote the translation from antichiral to chiral coordinates.
Introduce four independent scalar fluctuation amplitudes,
\begin{equation}
 \delta\W=\phi+\theta^4 c,\qquad
 \delta\bW=\mathcal T(\bar\phi+\bar\theta^4\bar c).
 \label{eq:unrestrictedscalar}
\end{equation}
Reduced chirality subsequently imposes $c=4P\bar\phi$ and
$\bar c=4P\phi$. The calculation uses eight odd coordinates and two
commuting plane-wave amplitudes of opposite momenta. Terms quadratic in these amplitudes give the Hessian.
For any numerator $J$ multiplying $Q^{-n}$, put $H=Q^2-Q_0^2$,
$Q_0=16fg$. Since $H$ vanishes on the background,
\begin{equation}
 \left[\frac{J}{Q^n}\right]_{(2)}
 =\frac1{Q_0^n}\left[J-\frac{nJH}{2Q_0^2}
                  +\frac{n(n+2)JH^2}{8Q_0^4}\right]_{(2)}.
 \label{eq:quadraticrootexpansion}
\end{equation}
To evaluate the mixed lowest-scalar entry, take $\phi$ and $\bar\phi$
with momenta $p_{\alpha\dot\alpha}$ and $-p_{\alpha\dot\alpha}$,
respectively, and set $c=\bar c=0$ temporarily. Write
$\ell=\sum_i\theta_i^\alpha p_{\alpha\dot\alpha}
\bar\theta^{\dot\alpha i}$, so that
$\mathcal T\bar\phi=e^{-2\ell}\bar\phi$. Let
$[\,\cdot\,]_*$ select the coefficient of
$\theta^4\bar\theta^4\phi\bar\phi$, with the two plane-wave factors
removed. Grassmann multiplication and the background tensor give
\begin{equation}
 \begin{gathered}
 \Theta^2=-2\theta^4,\qquad
 \bar\Theta^2=-2\bar\theta^4,\qquad
 [\Theta\bar\Theta\ell^2]_{\theta^4\bar\theta^4}=-4\Sigma,\\
 (\hK_0)_{\alpha\beta\dot\alpha\dot\beta}
       =8fg\,\delta_{\alpha\beta}\delta_{\dot\alpha\dot\beta},
 \qquad \hK_0^{vw}p_vp_w=16fg\Sigma,\qquad
 \Sigma=\omega^2+q_1^2-q_2^2+q_3^2.
 \end{gathered}
 \label{eq:scalargrassmanncontractions}
\end{equation}
The Kronecker symbols in this expression specify the components in the
frame \eqref{eq:componentbackground}; they are not invariant spinor
metrics. Tensor indices are raised with the antisymmetric spinor tensors
as in the definition of $\hK\cdot\hK$.

For $T_1$, put $J=2\W^2\bW^2$ and denote its terms of fluctuation
order $j$ by $J_j$. Similarly write $H=H_1+H_2+\cdots$. The terms needed
in \eqref{eq:quadraticrootexpansion} are
\begin{equation}
 \begin{aligned}
 J_0&=8f^2g^2\theta^4\bar\theta^4,\\
 J_1&=8fg^2\Theta\bar\theta^4\phi
       -8f^2g\theta^4\bar\Theta\bar\phi,\\
 (J_2)_{\phi\bar\phi}
    &=-8fg\Theta\bar\Theta e^{-2\ell}\phi\bar\phi,
       \qquad [J_2]_* =64fg\Sigma.
 \end{aligned}
 \label{eq:scalarT1numerator}
\end{equation}
Only the $2\ell^2$ term of the exponential contributes to the last
coefficient. To obtain the denominator terms, expand
$\hK=\hK_0+\hK_1+\hK_2$. Then
\begin{equation}
 H_1=2\hK_0\cdot\hK_1,\qquad
 H_2=\hK_1\cdot\hK_1+2\hK_0\cdot\hK_2.
 \label{eq:scalarHvariations}
\end{equation}
Let $H_\phi$ and $H_{\bar\phi}$ denote the coefficients of the two
lowest-scalar waves in $H_1$. For the wave of momentum $p$,
\begin{equation}
 \begin{aligned}
 D^i_\alpha\phi&=-2\bar\theta^{\dot\beta i}
                    p_{\alpha\dot\beta}\phi,\\
 D^i_\alpha D^j_\beta\phi
   &=4\bar\theta^{\dot\rho i}\bar\theta^{\dot\sigma j}
                p_{\alpha\dot\rho}p_{\beta\dot\sigma}\phi,
 \qquad
 \bar D_{\dot\alpha i}\bW_F
   =-g\frac{\partial\bar\Theta}{\partial\bar\theta^{\dot\alpha i}}.
 \end{aligned}
 \label{eq:scalarlinearspinorderivatives}
\end{equation}
Inserting these expressions into \eqref{eq:basis} gives
$[\bar\Theta\hK_0\cdot(X_\phi,Z_\phi,T_\phi)]_{\bar\theta^4}
=32fg^2\Sigma(1,-2i,8)$ at $\theta=0$; the subscript $\phi$
denotes the coefficient of that wave. Multiplication by twice the
coefficients $(-32,-32i,-4)$ in $\hK$ gives the components selected by
$J_1$:
\begin{equation}
 \begin{aligned}
 \left.[\bar\Theta H_\phi]_{\bar\theta^4}\right|_{\theta=0}
   &=(-2048-4096-2048)fg^2\Sigma=-8192fg^2\Sigma,\\
 \left.[\Theta H_{\bar\phi}]_{\theta^4}\right|_{\bar\theta=0}
   &=8192f^2g\Sigma.
 \end{aligned}
 \label{eq:scalarHlinearcontractions}
\end{equation}
The three contributions in the first line come from the $X$, $Z$ and
$T$ terms of \eqref{eq:basis}, with their coefficients in
$\hK$; the $Y$ term vanishes because one field is constant. The second
line follows from the barred calculation. The lowest component of
$(\hK_2)_{\phi\bar\phi}$ is $192p_vp_w\phi\bar\phi$:
$X$ contributes $64p_vp_w$ and $Y$ contributes $128p_vp_w$.
The linear scalar variations of $\hK$ have positive Grassmann degree,
so $\hK_1\cdot\hK_1$ has no lowest component. Consequently
\begin{equation}
 \begin{gathered}
 (H_2)_{\theta=\bar\theta=0}
        =6144fg\Sigma\phi\bar\phi,\\
 [J_1H_1]_*=131072f^3g^3\Sigma,
 \qquad [J_0H_2]_*=49152f^3g^3\Sigma,
 \qquad [J_0H_1^2]_*=0.
 \end{gathered}
 \label{eq:scalarT1denominator}
\end{equation}
The last equality holds because $J_0$ already contains all eight
Grassmann coordinates and $H_1$ has no lowest component. Substitution
in \eqref{eq:quadraticrootexpansion}, with $n=1$, gives
\begin{equation}
 \left[\frac{2\W^2\bW^2}{Q}\right]_*
  =\left(4-16-6\right)\Sigma=-18\Sigma.
 \label{eq:scalarT1sum}
\end{equation}
This calculation retains the two denominator contributions; expanding
only the numerator would instead give $4\Sigma$.

For $D_{33}$, the nilpotency of $\Theta$ gives $\W_F^3=\bW_F^3=0$.
The first variations are
\begin{equation}
 A:=(\W^3)_{(1)}=3f^2\Theta^2\phi=-6f^2\theta^4\phi,
 \qquad
 \bar B:=(\bW^3)_{(1)}=-6g^2\bar\theta^4\bar\phi.
 \label{eq:scalarcubicvariations}
\end{equation}
The antichiral translation in $\bar B$ adds no term because
$\bar\theta^4$ already contains all four barred coordinates. There is
no $c$ or $\bar c$ contribution: multiplication by
$\W_F^2$ or $\bW_F^2$ annihilates the corresponding highest scalar
component. Thus $K_{33}$ begins at second fluctuation order, and
\begin{equation}
 (D_{33})_{(2)}
   =\IF\!\left[\frac{\hK_0\cdot K_{33}[A,\bar B]}{(16fg)^3}\right].
 \label{eq:scalarD33quadratic}
\end{equation}
There are no denominator variations at this order. Applying
\eqref{eq:componentoperators} to $A,\bar B$ gives the following six
contractions, with the tensor definitions in \eqref{eq:mixedbasis}:
\begin{equation}
 \begin{array}{c|rrrrrr}
 B&X_L&X_R&Y&Z_L&Z_R&T\\\hline
 [\hK_0\cdot B]_* /(576f^3g^3\Sigma)
       &-1&-1&1&4i&-4i&-8\\
 \text{coefficient in }K_{33}&48&48&-128&16i&-16i&1
 \end{array}
 \label{eq:scalarD33contractions}
\end{equation}
Here $576=36\times16$: the product $A\bar B$ contributes $36f^2g^2$,
and contraction of the two momenta with $\hK_0$ gives $16fg\Sigma$.
For example, two derivatives on $A$ give $-p_vp_wA$, proving the first
entry. In $Z_L$, only the $2i\bar\theta\partial$ part of $D A$
can supply four unbarred coordinates; the sum over the two internal
indices gives $4ip_vp_w$. The opposite momentum in $Z_R$ reverses
this sign. For $T$, both unbarred derivatives must use their
$2i\bar\theta\partial$ parts. The Grassmann identity
\begin{equation}
 \bar\theta^A\bar\theta^B\partial_C\partial_D\bar\theta^4
   =(\delta^A_D\delta^B_C-\delta^A_C\delta^B_D)\bar\theta^4
 \label{eq:scalarbarcontraction}
\end{equation}
uses combined dotted and internal indices. After summing those internal
indices and symmetrizing the tensor indices, its two terms give
$(8-16)p_vp_w=-8p_vp_w$. This proves the last entry and hence
\begin{equation}
 \left[\frac{\hK\cdot K_{33}}{Q^3}\right]_*
  =\frac9{64}\left(-48-48-128-64-64-8\right)\Sigma
  =-\frac{405}{8}\Sigma.
 \label{eq:scalarD33sum}
\end{equation}

It remains to check the highest scalar components independently of
reduced chirality. For the $\phi c$ entry of $T_1$, the numerator
contributes $-8g^2\theta^4\bar\theta^4\phi c$.
The degree-two part of the $c$ coefficient in $H_1$ is
$256fg^2\Theta c$. Combining it with
$8fg^2\Theta\bar\theta^4\phi$ in $J_1$ gives
$[J_1H_1]_{\phi c}=-4096f^2g^4$. Therefore the two contributions cancel:
\begin{equation}
 (T_1)_{\phi c}/C_C=-\frac{g}{2f}+\frac{g}{2f}=0,
 \qquad
 (T_1)_{\bar\phi\bar c}/C_C=-\frac{f}{2g}+\frac{f}{2g}=0.
 \label{eq:scalarhighestcancellation}
\end{equation}
For $\phi\bar c$ and $c\bar\phi$, the numerator vanishes because it
contains more than four barred or unbarred coordinates; the same count
annihilates $J_1H_1$. The pairs $cc,c\bar c,\bar c\bar c$ have no
numerator contribution, since the square of a highest component and
its product with the electromagnetic background vanish. Their
$H_2$ terms have positive Grassmann degree and are annihilated by $J_0$.
For $\phi\phi$ or $\bar\phi\bar\phi$, the numerator lacks one set of
four coordinates, while every denominator contribution exceeds the
available degree in the other set. This argument can be made in chiral
or antichiral coordinates, respectively, since the total momentum is
zero. All these entries of $D_{33}$ vanish already by
\eqref{eq:scalarcubicvariations}, which contains only one lowest scalar
from each chirality.

The complete ten symmetric entries, divided by $C_C$, are therefore
\begin{equation}
 \begin{array}{c|cccc}
 \text{entry}&T_1&D_{33}&f_{\rm mod}&\Delta\\\hline
 \phi\bar\phi&-18\Sigma&-\frac{405}{8}\Sigma&\Sigma&19\Sigma\\
 \phi\phi,\ \phi c,\ \phi\bar c&0&0&0&0\\
 cc,\ c\bar\phi,\ c\bar c&0&0&0&0\\
 \bar\phi\bar\phi,\ \bar\phi\bar c,\ \bar c\bar c&0&0&0&0
 \end{array}
 \label{eq:scalarhessiandata}
\end{equation}
All background parameters and momentum components were retained in the
contractions. The factors of $f,g$ cancel separately, and the only
momentum dependence is the quadratic polynomial $\Sigma$.
The chosen coefficient in \eqref{eq:additionalinteractions} follows from
$-18-(405/8)a=1$, giving $a=-152/405$.

The Maxwell coefficient after imposing reduced chirality is
$2C_CP\phi\bar\phi$, fixing $z=\sqrt{2C_C}\phi$. Hence
\begin{equation}
 (f_{\rm mod})^{(2)}_z=\int\dd^4x\,\frac{T_F^{\mu\nu}}{2r}
                  \partial_\mu z\partial_\nu\bar z,\qquad
 (\Delta)^{(2)}_z=19(f_{\rm mod})^{(2)}_z.
 \label{eq:modifiedscalarinput}
\end{equation}
The minimal interaction $T_1$ instead has coefficient $-9T_F^{\mu\nu}/r$.
The additional primary density thus changes the scalar Hessian while leaving
the truncation to the $\mathcal N=1$ vector multiplet unchanged.

\subsection{Closure under the recursion}
The scalar first variations vanish on the background. Reflection
excludes mixed scalar--photon variations, and $SU(2)_R$ excludes
scalar--auxiliary-triplet variations, including for the independent scalar components of the unconstrained chiral superfields. Thus \eqref{eq:scalarhessiandata} gives the complete scalar Hessian.
For scalar component variations $h$, the second variation of a pairing is
\begin{equation}
 \begin{aligned}
 \dd^2\PP(F,G)[h,h]={}&
 \langle\dd^2F_{\W}[h,h],G_{\W,0}\rangle_C
 +2\langle\dd F_{\W}[h],\dd G_{\W}[h]\rangle_C\\
 &+\langle F_{\W,0},\dd^2G_{\W}[h,h]\rangle_C,
 \qquad\langle A,B\rangle_C=\IC[AB].
 \end{aligned}
 \label{eq:scalarpairvariation}
\end{equation}Assume the scalar Hessians of $F,G$ are constant multiples of
$\Sigma\phi\bar\phi$, including zero. The linear variation of their chiral
functional derivatives is then confined to the $\theta^4$ component: the
lowest scalar variation pairs with $\theta^4$ under the chiral measure. The
middle term vanishes by $(\theta^4)^2=0$, as does its barred counterpart.

On this background, $U(1)_R$, $SU(2)_R$ and fermion parity restrict
the chiral functional derivatives evaluated on the background to the electromagnetic components:
\begin{equation}
 F_{\W,0}=b_F\Theta,\qquad G_{\W,0}=b_G\Theta.
 \label{eq:backgroundresponse}
\end{equation}
Here $b_F,b_G$ denote their coefficient functions, not the magnetic
amplitude in \eqref{eq:componentbackground}.
Let $H_F[h,h]=\dd^2F[h,h]$ be an independent-chiral scalar Hessian,
with the wave directions and their momentum held fixed. Since
$\W_0=-f\Theta$, a fixed variation $\delta\W=b_G\Theta$ is
$\delta f=-b_G$. Symmetry of the functional third variation then gives
\begin{equation}
 \left\langle\dd^2F_{\W}[h,h],G_{\W,0}\right\rangle_C
   =-b_G\,\partial_f H_F[h,h],\qquad
 \dd^2\PP(F,G)[h,h]
   =-b_G\partial_fH_F-b_F\partial_fH_G.
 \label{eq:explicitradialclosure}
\end{equation}
No derivative of $b_F$ or $b_G$ occurs: the third variation is evaluated
in a fixed background direction. The antichiral pairing gives the
corresponding $\partial_g$ derivatives, with the sign fixed by
$\bW_0=g\bar\Theta$. Every derivative vanishes because the full
independent-chiral Hessians in \eqref{eq:scalarhessiandata} are
independent of $f$ and $g$ separately. Together with the vanishing
middle term, this proves closure of the scalar Hessian under the
pairing. The chiral and antichiral variations have remained independent
throughout this calculation.

The independent-chiral scalar Hessian of $\Delta$ has the same
form, by \eqref{eq:scalarhessiandata}. Thus each prescribed $C_n$ contributes
a Hessian of this form at the next order.

Every pairing that enters the recursion therefore has zero scalar Hessian if
all lower scalar Hessians have the stated form. The phase average
commutes with the expansion when the background and fluctuations are both
transformed. The undetermined scalar coefficient at each order is
phase-invariant and fixed by the prescribed phase-invariant part. Induction gives
\begin{equation}
 \boxed{(F_n)^{(2)}_z=0,\qquad
 (\widetilde F_n)^{(2)}_z=C_n^{(2)}
 =[t^n]d(t)\,(\Delta)^{(2)}_z\quad(n\geq2).}
 \label{eq:scalarclosure}
\end{equation}
There is consequently no higher-derivative scalar quadratic operator
at any formal order on \eqref{eq:constantbackground}. This conclusion
is specific to the stated backgrounds and prescribed coefficients.

\subsection{Analytic continuation of the quadratic action}
For the canonical completion the transverse spatial coefficient
vanishes at
\begin{equation}
 \gamma_* =\operatorname{arsinh}2=\log(2+\sqrt5)
 \label{eq:canonicalthreshold}
\end{equation}
and becomes negative beyond it. For the action, summing the
known quadratic coefficients gives
$\beta(t)=(t+19d(t))/2=t/(\sqrt{1+t^2}+1)$, and hence
\eqref{eq:scalardispersion}. The real positive branch specifies an
analytic continuation at every finite $t\geq0$, with positive kinetic
coefficients and second-order field equations.
This second-order scalar operator determines the characteristic cone on
these backgrounds. The characteristics of the coupled system on other
backgrounds require a separate analysis.

\section{Derivation of the special-supersymmetry variations}
\label{app:svariation}
We derive the variations of the derivatives and bilinear products used in
\eqref{eq:jetvariations} and \eqref{eq:basisvariation}.
Use the polarized derivatives of \eqref{eq:polarization}, with
$p_y=p$ in this appendix, and put
$F^{ij}=D^iD^j\W=-F^{ji}$ and
$G_{ij}=\bar D_i\bar D_j\bW=-G_{ji}$.
The derivative algebra is $\{D^i,\bar D_j\}=-2i\delta^i{}_jp$.
The parameters $\chi^i,\bar\chi_i$ and first descendants $F^i,G_i$
are odd; the variation including its parameters is even.

For pure special supersymmetry, the flat Killing constraints
\cite{K2008} give
\begin{equation}
 \begin{aligned}
 D^i\omega_y{}^\beta&=-\chi^iy^\beta,&
 D^i\Lambda_j{}^k&=-2\delta_j{}^i\chi^k+\delta_j{}^k\chi^i,\\
 \bar D_i\bar\omega_{\bar y}{}^{\dot\beta}
       &=-\bar\chi_i\bar y^{\dot\beta},&
 \bar D_i\Lambda_j{}^k&=2\delta_i{}^k\bar\chi_j-\delta_j{}^k\bar\chi_i.
 \end{aligned}
 \label{eq:wardparameterderivatives}
\end{equation}
Chirality sets the crossed derivatives of $\sigma,\bar\sigma$ and
of the Lorentz parameters to zero. The odd parameters are constant,
and $p$ annihilates the parameter functions. For either fixed
internal index $h$ (not summed), the Jacobi identity yields
\begin{equation}
 [\xi,p]=\frac{i}{2}\bigl(
 \{[\xi,D^h],\bar D_h\}+\{D^h,[\xi,\bar D_h]\}\bigr),\qquad
 [\xi,p]_{\rm inh}=-i\bar\chi_iD^i-i\chi^i\bar D_i.
 \label{eq:wardvectorcommutator}
\end{equation}
Indeed, the coefficients of $D^k,\bar D_k$ inside the parentheses
are $-\delta_k{}^h\bar\chi_h-\bar D_h\Lambda_k{}^h=-2\bar\chi_k$
and $-\delta_h{}^k\chi^h+D^h\Lambda_h{}^k=-2\chi^k$.

Differentiating \eqref{eq:Wprimary} once gives
\begin{equation}
 \delta F^j=-\xi F^j+\omega_y{}^\beta D^j_\beta\W
 -(2\sigma+\bar\sigma)F^j-\Lambda_k{}^jF^k-2\chi^j\W.
 \label{eq:wardfirstdescendant}
\end{equation}
A further $D^i$ derivative has four inhomogeneous contributions:
\begin{equation}
 \begin{aligned}
 (D^i\omega_y{}^\beta)D^j_\beta\W&=-\chi^iF^j,&
 -(D^i\Lambda_k{}^j)F^k&=-\chi^iF^j+2\chi^jF^i,\\
 -D^i(2\sigma+\bar\sigma)F^j&=-2\chi^iF^j,&
 -2D^i(\chi^j\W)&=2\chi^jF^i.
 \end{aligned}
 \label{eq:wardsecondterms}
\end{equation}
Their sum is $\AS(F^{ij})=4(\chi^jF^i-\chi^iF^j)$.
For the mixed derivative, chirality gives
$\bar D_jF^i=-2i\delta_j{}^ip\W$, so
\begin{equation}
 \AS(pF^i)=-i\bar\chi_jD^jF^i-i\chi^j\bar D_jF^i-2\chi^ip\W
          =-i\bar\chi_jF^{ji}-4\chi^ip\W.
 \label{eq:wardmixeddescendant}
\end{equation}
The two contributions to $-4\chi^ip\W$ come from the derivative
anticommutator and the last term of \eqref{eq:wardfirstdescendant}.
The barred formulas follow by conjugation. One more spacetime
derivative gives $\AS(p^2\W)=-2i\bar\chi_i pF^i$ and its conjugate.

For the products, $X=(p^2\W)\bW+\W p^2\bW$ and
$Y=(p\W)(p\bW)$ immediately give the first two entries of
\eqref{eq:basisvariation}. For $Z=(pF^i)G_i-F^i(pG_i)$, moving
all odd parameters to the left gives
\begin{equation}
 \begin{aligned}
 \AS(Z)={}&-4\chi^i(p\W)G_i-i\bar\chi_jF^{ji}G_i
          +2\bar\chi_i(pF^i)\bW\\
 &+2\chi^i\W pG_i-4\bar\chi_iF^ip\bW-i\chi^jF^iG_{ji}
 =2U_S-4V_S-iM_S.
 \end{aligned}
 \label{eq:wardZ}
\end{equation}
In particular, $(pF^i)(-2\bar\chi_i\bW)=+2\bar\chi_i(pF^i)\bW$.
Antisymmetry gives
\[
 \AS(F^{ij}G_{ij})=-8(\chi^iF^jG_{ij}+\bar\chi_iF^{ij}G_j)=-8M_S.
\]
These results establish the coefficient equations in
Section~\ref{sec:primary} without a Bianchi identity or field equation.

We now reserve $p$ for the chiral weight and write the spacetime
derivative as $p_y$. For $A=\W^p$, differentiating the weight term
replaces $-2\chi^i\W$ by $-2p\chi^iA$. The parameter derivatives
are unchanged, giving
\begin{equation}
 \begin{aligned}
 \AS(D^iA)&=-2p\chi^iA,& \AS(p_yA)&=-i\bar\chi_iD^iA,\\
 \AS(D^iD^jA)&=2(p+1)(\chi^jD^iA-\chi^iD^jA),\\
 \AS(p_yD^iA)&=-2(p+1)\chi^ip_yA-i\bar\chi_jD^jD^iA,\\
 \AS(p_y^2A)&=-2i\bar\chi_i p_yD^iA.
 \end{aligned}
 \label{eq:weighteddescendants}
\end{equation}
The conjugate formulas have weight $q$ for $\bW^q$.
Applying these identities to the six products in
\eqref{eq:mixedbasis} gives \eqref{eq:mixedwardmatrix}.

\section*{Acknowledgments}
\emph{Use of AI tools.} OpenAI GPT-6 Astra and Anthropic Claude Opus~5.5
were used to assist with derivations, symbolic verification codes and
editing of the texts. The tools were given the
manuscript, calculation notes and specific derivation and verification
tasks. Symbolic calculations were also used to check the component identities
and quadratic propagation formulas. The derivations are given in the paper. The author takes full
responsibility for the content of this paper.

\bibliographystyle{unsrtnat}
\bibliography{n2_modmax_references}

@article{BLST2020,
  author = {Bandos, Igor and Lechner, Kurt and Sorokin, Dmitri and Townsend, Paul K.},
  title = {A non-linear duality-invariant conformal extension of {Maxwell}'s equations},
  year = {2020},
  eprint = {2007.09092},
  journal = {Phys. Rev. D},
  volume = {102},
  pages = {121703},
  doi = {10.1103/PhysRevD.102.121703},
  note = {\href{https://arxiv.org/abs/2007.09092}{arXiv:2007.09092}}
}

@article{BLST2021,
  author = {Bandos, Igor and Lechner, Kurt and Sorokin, Dmitri and Townsend, Paul K.},
  title = {{ModMax} meets {Susy}},
  year = {2021},
  eprint = {2106.07547},
  journal = {JHEP}, volume = {10}, pages = {031},
  doi = {10.1007/JHEP10(2021)031},
  note = {\href{https://arxiv.org/abs/2106.07547}{arXiv:2106.07547}}
}

@article{KT2000,
  author = {Kuzenko, Sergei M. and Theisen, Stefan},
  title = {Supersymmetric duality rotations},
  year = {2000},
  eprint = {hep-th/0001068},
  journal = {JHEP},
  volume = {03},
  pages = {034},
  doi = {10.1088/1126-6708/2000/03/034},
  note = {\href{https://arxiv.org/abs/hep-th/0001068}{arXiv:hep-th/0001068}}
}

@article{K2013,
  author = {Kuzenko, Sergei M.},
  title = {Duality rotations in supersymmetric nonlinear electrodynamics revisited},
  year = {2013},
  eprint = {1301.5194},
  journal = {JHEP},
  volume = {03},
  pages = {153},
  doi = {10.1007/JHEP03(2013)153},
  note = {\href{https://arxiv.org/abs/1301.5194}{arXiv:1301.5194}}
}

@article{K2021,
  author = {Kuzenko, Sergei M.},
  title = {Superconformal duality-invariant models and {$\mathcal N=4$} {SYM} effective action},
  year = {2021},
  eprint = {2106.07173},
  journal = {JHEP},
  volume = {09},
  pages = {180},
  doi = {10.1007/JHEP09(2021)180},
  note = {\href{https://arxiv.org/abs/2106.07173}{arXiv:2106.07173}}
}

@article{KR2021,
  author = {Kuzenko, Sergei M. and Raptakis, Emmanouil S. N.},
  title = {Duality-invariant (super)conformal higher-spin models},
  year = {2021},
  eprint = {2107.02001},
  journal = {Phys. Rev. D},
  volume = {104},
  pages = {125003},
  doi = {10.1103/PhysRevD.104.125003},
  note = {\href{https://arxiv.org/abs/2107.02001}{arXiv:2107.02001}}
}

@article{K2008,
  author = {Kuzenko, Sergei M.},
  title = {On superconformal projective hypermultiplets},
  year = {2007},
  eprint = {0710.1479},
  journal = {JHEP},
  volume = {12},
  pages = {010},
  doi = {10.1088/1126-6708/2007/12/010},
  note = {\href{https://arxiv.org/abs/0710.1479}{arXiv:0710.1479}}
}

@article{FST2022,
  author = {Ferko, Christian and Smith, Liam and Tartaglino-Mazzucchelli, Gabriele},
  title = {On current-squared flows and {ModMax} theories},
  year = {2022},
  eprint = {2203.01085},
  journal = {SciPost Phys.},
  volume = {13},
  pages = {012},
  doi = {10.21468/SciPostPhys.13.2.012},
  note = {\href{https://arxiv.org/abs/2203.01085}{arXiv:2203.01085}}
}

@article{FST2023,
  author = {Ferko, Christian and Smith, Liam and Tartaglino-Mazzucchelli, Gabriele},
  title = {Stress tensor flows, birefringence in non-linear electrodynamics, and supersymmetry},
  year = {2023},
  eprint = {2301.10411},
  journal = {SciPost Phys.},
  volume = {15},
  pages = {198},
  doi = {10.21468/SciPostPhys.15.5.198},
  note = {\href{https://arxiv.org/abs/2301.10411}{arXiv:2301.10411}}
}

@article{BK2010,
  author = {Butter, Daniel and Kuzenko, Sergei M.},
  title = {{$\mathcal N=2$} supergravity and supercurrents},
  year = {2010},
  eprint = {1011.0339},
  journal = {JHEP},
  volume = {12},
  pages = {080},
  doi = {10.1007/JHEP12(2010)080},
  note = {\href{https://arxiv.org/abs/1011.0339}{arXiv:1011.0339}}
}

@article{KR2023,
  author = {Kuzenko, Sergei M. and Raptakis, Emmanouil S. N.},
  title = {On higher-spin {$\mathcal N=2$} supercurrent multiplets},
  year = {2023},
  eprint = {2301.09386},
  journal = {JHEP},
  volume = {05},
  pages = {056},
  doi = {10.1007/JHEP05(2023)056},
  note = {\href{https://arxiv.org/abs/2301.09386}{arXiv:2301.09386}}
}

@article{KS2023,
  author = {Kuzenko, Sergei M. and Stirling, Jake C.},
  title = {New duality-invariant models for nonlinear supersymmetric electrodynamics},
  year = {2023},
  eprint = {2308.07113},
  journal = {JHEP},
  volume = {12},
  pages = {041},
  doi = {10.1007/JHEP12(2023)041},
  note = {\href{https://arxiv.org/abs/2308.07113}{arXiv:2308.07113}}
}

@article{K2026,
  author = {Kuzenko, Sergei M.},
  title = {Nonlinear self-duality for arbitrary spin, superspin, and supersymmetry types},
  year = {2026},
  eprint = {2602.04336},
  journal = {Universe},
  volume = {12},
  number = {6},
  pages = {179},
  doi = {10.3390/universe12060179},
  note = {\href{https://arxiv.org/abs/2602.04336}{arXiv:2602.04336}}
}

@article{Shore2007,
  author = {Shore, Graham M.},
  title = {Superluminality and {UV} completion},
  year = {2007},
  eprint = {hep-th/0701185},
  journal = {Nucl. Phys. B},
  volume = {778},
  pages = {219--258},
  doi = {10.1016/j.nuclphysb.2007.03.034},
  note = {\href{https://arxiv.org/abs/hep-th/0701185}{arXiv:hep-th/0701185}}
}

@article{KTReview2001,
  author = {Kuzenko, Sergei M. and Theisen, Stefan},
  title = {Nonlinear Self-Duality and Supersymmetry},
  year = {2001},
  eprint = {hep-th/0007231},
  doi = {10.1002/1521-3978(200102)49:1/3<273::aid-prop273>3.0.co;2-0},
  journal = {Fortsch. Phys.},
  volume = {49},
  pages = {273-309},
  note = {\href{https://arxiv.org/abs/hep-th/0007231}{arXiv:hep-th/0007231}}
}

@article{KM2003,
  author = {Kuzenko, Sergei M. and McCarthy, Shane A.},
  title = {Nonlinear Self-Duality and Supergravity},
  year = {2003},
  eprint = {hep-th/0212039},
  journal = {JHEP},
  volume = {02},
  pages = {038},
  doi = {10.1088/1126-6708/2003/02/038},
  note = {\href{https://arxiv.org/abs/hep-th/0212039}{arXiv:hep-th/0212039}}
}

@article{KM2005,
  author = {Kuzenko, S. M. and McCarthy, S. A.},
  title = {On the component structure of {$\mathcal N=1$} supersymmetric nonlinear electrodynamics},
  year = {2005},
  eprint = {hep-th/0501172},
  doi = {10.1088/1126-6708/2005/05/012},
  journal = {JHEP},
  volume = {05},
  pages = {012},
  note = {\href{https://arxiv.org/abs/hep-th/0501172}{arXiv:hep-th/0501172}}
}

@article{KTy2011,
  author = {Kuzenko, Sergei M. and Tyler, Simon J.},
  title = {On the {Goldstino} actions and their symmetries},
  year = {2011},
  eprint = {1102.3043},
  journal = {JHEP},
  volume = {05},
  pages = {055},
  doi = {10.1007/JHEP05(2011)055},
  note = {\href{https://arxiv.org/abs/1102.3043}{arXiv:1102.3043}}
}

@article{KMcA2023,
  author = {Kuzenko, Sergei M. and McArthur, I. N.},
  title = {A supersymmetric nonlinear sigma model analogue of the {ModMax} theory},
  year = {2023},
  eprint = {2303.15139},
  doi = {10.1007/JHEP05(2023)127},
  journal = {JHEP},
  volume = {05},
  pages = {127},
  note = {\href{https://arxiv.org/abs/2303.15139}{arXiv:2303.15139}}
}

@article{FKST2023,
  author = {Ferko, Christian and Kuzenko, Sergei M. and Smith, Liam and Tartaglino-Mazzucchelli, Gabriele},
  title = {Duality-Invariant Non-linear Electrodynamics and Stress Tensor Flows},
  year = {2023},
  eprint = {2309.04253},
  doi = {10.1103/physrevd.108.106021},
  journal = {Phys. Rev. D},
  volume = {108},
  pages = {106021},
  note = {\href{https://arxiv.org/abs/2309.04253}{arXiv:2309.04253}}
}

@article{KR2024,
  author = {Kuzenko, Sergei M. and Raptakis, Emmanouil S. N.},
  title = {Higher-derivative deformations of the {ModMax} theory},
  year = {2024},
  eprint = {2404.09108},
  journal = {JHEP},
  volume = {06},
  pages = {162},
  doi = {10.1007/JHEP06(2024)162},
  note = {\href{https://arxiv.org/abs/2404.09108}{arXiv:2404.09108}}
}

@article{GR1995,
  author = {Gibbons, G W and Rasheed, D A},
  title = {Electric-Magnetic Duality Rotations in Non-Linear Electrodynamics},
  year = {1995},
  eprint = {hep-th/9506035},
  doi = {10.1016/0550-3213(95)00409-l},
  journal = {Nucl. Phys. B},
  volume = {454},
  pages = {185-206},
  note = {\href{https://arxiv.org/abs/hep-th/9506035}{arXiv:hep-th/9506035}}
}

@misc{IZ2002,
  author = {Ivanov, E. A. and Zupnik, B. M.},
  title = {New representation for {L}agrangians of self-dual nonlinear electrodynamics},
  year = {2002},
  eprint = {hep-th/0202203},
  note = {\href{https://arxiv.org/abs/hep-th/0202203}{arXiv:hep-th/0202203}}
}

@article{IZ2013,
  author = {Ivanov, E. A. and Zupnik, B. M.},
  title = {Bispinor Auxiliary Fields in Duality-Invariant Electrodynamics Revisited},
  year = {2013},
  eprint = {1212.6637},
  doi = {10.1103/physrevd.87.065023},
  journal = {Phys. Rev. D},
  volume = {87},
  pages = {065023},
  note = {\href{https://arxiv.org/abs/1212.6637}{arXiv:1212.6637}}
}

@article{BG1997,
  author = {Bagger, Jonathan and Galperin, Alexander},
  title = {New {Goldstone} multiplet for partially broken supersymmetry},
  year = {1997},
  eprint = {hep-th/9608177},
  doi = {10.1103/physrevd.55.1091},
  journal = {Phys. Rev. D},
  volume = {55},
  pages = {1091-1098},
  note = {\href{https://arxiv.org/abs/hep-th/9608177}{arXiv:hep-th/9608177}}
}

@article{Ketov1999,
  author = {Ketov, Sergei V.},
  title = {A manifestly {$\mathcal N=2$} supersymmetric {Born--Infeld} action},
  year = {1999},
  eprint = {hep-th/9809121},
  doi = {10.1142/s0217732399000559},
  journal = {Mod. Phys. Lett. A},
  volume = {14},
  pages = {501-510},
  note = {\href{https://arxiv.org/abs/hep-th/9809121}{arXiv:hep-th/9809121}}
}

@article{BIK2001,
  author = {Bellucci, S. and Ivanov, E. and Krivonos, S.},
  title = {Towards the complete {$\mathcal N=2$} superfield {Born--Infeld} action with partially broken {$\mathcal N=4$} supersymmetry},
  year = {2001},
  eprint = {hep-th/0101195},
  journal = {Phys. Rev. D},
  volume = {64},
  pages = {025014},
  doi = {10.1103/PhysRevD.64.025014},
  note = {\href{https://arxiv.org/abs/hep-th/0101195}{arXiv:hep-th/0101195}}
}

@article{BKSS2013,
  author = {Bellucci, S. and Krivonos, S. and Shcherbakov, A. and Sutulin, A.},
  title = {On the road to {$\mathcal N=2$} supersymmetric {Born--Infeld} action},
  year = {2013},
  eprint = {1212.1902},
  doi = {10.1016/j.physletb.2013.03.035},
  journal = {Phys. Lett. B},
  volume = {721},
  pages = {353-357},
  note = {\href{https://arxiv.org/abs/1212.1902}{arXiv:1212.1902}}
}

@article{Sorokin2022,
  author = {Sorokin, Dmitri P.},
  title = {Introductory Notes on Non-linear Electrodynamics and its Applications},
  year = {2022},
  eprint = {2112.12118},
  doi = {10.1002/prop.202200092},
  journal = {Fortsch. Phys.},
  volume = {70},
  pages = {2200092},
  note = {\href{https://arxiv.org/abs/2112.12118}{arXiv:2112.12118}}
}

@article{FSST2022,
  author = {Ferko, Christian and Sfondrini, Alessandro and Smith, Liam and Tartaglino-Mazzucchelli, Gabriele},
  title = {{Root-$T\overline{T}$} Deformations in Two-Dimensional Quantum Field Theories},
  year = {2022},
  eprint = {2206.10515},
  journal = {Phys. Rev. Lett.},
  volume = {129},
  pages = {201604},
  doi = {10.1103/PhysRevLett.129.201604},
  note = {\href{https://arxiv.org/abs/2206.10515}{arXiv:2206.10515}}
}

@article{BHO2024,
  author = {Babaei-Aghbolagh, H. and He, Song and Ouyang, Hao},
  title = {Generalized {$T\overline{T}$-like} Deformations in Duality-Invariant Nonlinear Electrodynamic Theories},
  year = {2024},
  eprint = {2407.03698},
  doi = {10.1007/jhep09(2024)137},
  journal = {JHEP},
  volume = {09},
  pages = {137},
  note = {\href{https://arxiv.org/abs/2407.03698}{arXiv:2407.03698}}
}

@article{RT2023,
  author = {Russo, Jorge G. and Townsend, Paul K.},
  title = {Nonlinear Electrodynamics without Birefringence},
  year = {2023},
  eprint = {2211.10689},
  doi = {10.1007/JHEP01(2023)039},
  journal = {JHEP},
  volume = {01},
  pages = {039},
  note = {\href{https://arxiv.org/abs/2211.10689}{arXiv:2211.10689}}
}

@article{RT2024a,
  author = {Russo, Jorge G. and Townsend, Paul K.},
  title = {Causal Self-Dual Electrodynamics},
  year = {2024},
  eprint = {2401.06707},
  journal = {Phys. Rev. D},
  volume = {109},
  pages = {105023},
  doi = {10.1103/PhysRevD.109.105023},
  note = {\href{https://arxiv.org/abs/2401.06707}{arXiv:2401.06707}}
}

@article{RT2024b,
  author = {Russo, Jorge G. and Townsend, Paul K.},
  title = {Causality and Energy Conditions in Nonlinear Electrodynamics},
  year = {2024},
  eprint = {2404.09994},
  doi = {10.1007/jhep06(2024)191},
  journal = {JHEP},
  volume = {06},
  pages = {191},
  note = {\href{https://arxiv.org/abs/2404.09994}{arXiv:2404.09994}}
}

@article{LMSS2023,
  author = {Lechner, Kurt and Marchetti, Pieralberto and Sainaghi, Andrea and Sorokin, Dmitri P.},
  title = {Maximally symmetric nonlinear extension of electrodynamics and charged particles},
  year = {2022},
  eprint = {2206.04657},
  doi = {10.1103/physrevd.106.016009},
  journal = {Phys. Rev. D},
  volume = {106},
  pages = {016009},
  note = {\href{https://arxiv.org/abs/2206.04657}{arXiv:2206.04657}}
}

@article{CCDGO2026,
  author = {Canfora, Fabrizio and Corral, Crist{\'o}bal and Diez, Borja and Guajardo, Luis and Oliva, Julio},
  title = {Self-dual instantons and gravitating dyons in non-{A}belian {ModMax} theory},
  year = {2026},
  eprint = {2511.15437},
  doi = {10.1007/JHEP03(2026)057},
  journal = {JHEP},
  volume = {03},
  pages = {057},
  note = {\href{https://arxiv.org/abs/2511.15437}{arXiv:2511.15437}}
}

@misc{DG2026,
  author = {Diez, Borja and Guajardo, Luis},
  title = {Merons beyond {Yang--Mills} theory},
  year = {2026},
  eprint = {2604.14046},
  note = {\href{https://arxiv.org/abs/2604.14046}{arXiv:2604.14046}}
}

@article{BMV2008,
 author={Babichev, Eugeny and Mukhanov, Viatcheslav and Vikman, Alexander},
 title={{k-Essence}, superluminal propagation, causality and emergent geometry},
 journal={JHEP}, volume={02}, pages={101}, year={2008},
 eprint={0708.0561}, doi={10.1088/1126-6708/2008/02/101},
  note = {\href{https://arxiv.org/abs/0708.0561}{arXiv:0708.0561}}
}

@article{ACGR2015,
 author={Abalos, Fernando and Carrasco, Federico and Goulart, {\'E}rico and Reula, Oscar},
 title={Nonlinear electrodynamics as a symmetric hyperbolic system},
 journal={Phys. Rev. D}, volume={92}, pages={084024}, year={2015},
 eprint={1507.02262}, doi={10.1103/PhysRevD.92.084024},
  note = {\href{https://arxiv.org/abs/1507.02262}{arXiv:1507.02262}}
}

@article{Reall2021,
 author={Reall, Harvey S.},
 title={Causality in gravitational theories with second order equations of motion},
 journal={Phys. Rev. D}, volume={103}, pages={084027}, year={2021},
 eprint={2101.11623}, doi={10.1103/PhysRevD.103.084027},
  note = {\href{https://arxiv.org/abs/2101.11623}{arXiv:2101.11623}}
}

@article{CGdRPT2022,
 author={Carrillo Gonz{\'a}lez, Mariana and de Rham, Claudia and Pozsgay, Victor and Tolley, Andrew J.},
 title={Causal effective field theories},
 journal={Phys. Rev. D}, volume={106}, pages={105018}, year={2022},
 eprint={2207.03491}, doi={10.1103/PhysRevD.106.105018},
  note = {\href{https://arxiv.org/abs/2207.03491}{arXiv:2207.03491}}
}

@article{BSTW2019,
  author = {Baggio, Marco and Sfondrini, Alessandro and Tartaglino-Mazzucchelli, Gabriele and Walsh, Harriet},
  title = {On {$T\bar T$} deformations and supersymmetry},
  journal = {JHEP}, volume = {06}, year = {2019}, pages = {063},
  doi = {10.1007/JHEP06(2019)063}, eprint = {1811.00533},
  archivePrefix = {arXiv}, primaryClass = {hep-th},
  note = {\href{https://arxiv.org/abs/1811.00533}{arXiv:1811.00533}}
}

@article{CFSST2020,
  author = {Chang, Chih-Kai and Ferko, Christian and Sethi, Savdeep and Sfondrini, Alessandro and Tartaglino-Mazzucchelli, Gabriele},
  title = {{$T\bar T$} flows and (2,2) supersymmetry},
  journal = {Phys. Rev. D}, volume = {101}, year = {2020}, pages = {026008},
  doi = {10.1103/PhysRevD.101.026008}, eprint = {1906.00467},
  archivePrefix = {arXiv}, primaryClass = {hep-th},
  note = {\href{https://arxiv.org/abs/1906.00467}{arXiv:1906.00467}}
}

@article{IZ2004,
  author = {Ivanov, E. A. and Zupnik, B. M.},
  title = {New approach to nonlinear electrodynamics: Dualities as symmetries of interaction},
  year = {2004},
  journal = {Phys. Atom. Nucl.},
  volume = {67},
  pages = {2188--2199},
  doi = {10.1134/1.1842299},
  eprint = {hep-th/0303192},
  note = {\href{https://arxiv.org/abs/hep-th/0303192}{arXiv:hep-th/0303192}}
}

@article{ILZ2013,
  author = {Ivanov, Evgeny and Lechtenfeld, Olaf and Zupnik, Boris},
  title = {Auxiliary superfields in {$\mathcal N=1$} supersymmetric self-dual electrodynamics},
  year = {2013},
  journal = {JHEP},
  volume = {05},
  pages = {133},
  doi = {10.1007/JHEP05(2013)133},
  eprint = {1303.5962},
  note = {\href{https://arxiv.org/abs/1303.5962}{arXiv:1303.5962}}
}

@article{IZ2014,
  author = {Ivanov, E. A. and Zupnik, B. M.},
  title = {Self-dual {$\mathcal N=2$} {Born--Infeld} theory through auxiliary superfields},
  year = {2014},
  journal = {JHEP},
  volume = {05},
  pages = {061},
  doi = {10.1007/JHEP05(2014)061},
  eprint = {1312.5687},
  note = {\href{https://arxiv.org/abs/1312.5687}{arXiv:1312.5687}}
}

@article{BIZ2022,
  author = {Buchbinder, Ioseph and Ivanov, Evgeny and Zaigraev, Nikita},
  title = {Off-shell cubic hypermultiplet couplings to {$\mathcal N=2$} higher spin gauge superfields},
  year = {2022},
  journal = {JHEP},
  volume = {05},
  pages = {104},
  doi = {10.1007/JHEP05(2022)104},
  eprint = {2202.08196},
  note = {\href{https://arxiv.org/abs/2202.08196}{arXiv:2202.08196}}
}

@article{BIZ2023,
  author = {Buchbinder, Ioseph and Ivanov, Evgeny and Zaigraev, Nikita},
  title = {{$\mathcal N=2$} higher spins: superfield equations of motion, the hypermultiplet supercurrents, and the component structure},
  year = {2023},
  journal = {JHEP},
  volume = {03},
  pages = {036},
  doi = {10.1007/JHEP03(2023)036},
  eprint = {2212.14114},
  note = {\href{https://arxiv.org/abs/2212.14114}{arXiv:2212.14114}}
}

@article{BBYM2022,
  author = {Babaei-Aghbolagh, H. and Babaei Velni, Komeil and Mahdavian Yekta, Davood and Mohammadzadeh, H.},
  title = {Emergence of non-linear electrodynamic theories from {$T\bar T$}-like deformations},
  year = {2022},
  journal = {Phys. Lett. B},
  volume = {829},
  pages = {137079},
  doi = {10.1016/j.physletb.2022.137079},
  eprint = {2202.11156},
  note = {\href{https://arxiv.org/abs/2202.11156}{arXiv:2202.11156}}
}

@article{Z2026,
  author = {Zaigraev, Nikita},
  title = {Structure of {$\mathcal N=2$} superfield higher-spin abelian cubic interactions},
  year = {2026},
  journal = {JHEP},
  volume = {09},
  pages = {181},
  doi = {10.1007/JHEP09(2026)181},
  eprint = {2605.27206},
  note = {\href{https://arxiv.org/abs/2605.27206}{arXiv:2605.27206}}
}

\end{document}